\documentclass[12pt,a4paper]{article}
\pdfoutput=1
\usepackage{afterpage}
\usepackage[a4paper]{geometry}
\usepackage{float}
\usepackage[stable]{footmisc}
\usepackage[utf8x]{inputenc}
\usepackage{a4wide}
\usepackage{amsmath,amssymb,amstext,amsthm,amssymb}
\usepackage{amsfonts}
\usepackage{subcaption}
\usepackage{setspace}
\usepackage{bbold}
\usepackage{bbm}
\usepackage{amstext}
\usepackage{array}
\usepackage{hyperref}
\usepackage{overpic}
\usepackage{bbm}
\usepackage[nosort]{cite}
\usepackage{lipsum}
\usepackage{color}
\usepackage{enumitem}
\usepackage{cancel}

\graphicspath{{figures/}}

\newcommand{\diff}[0]{\text{d}}
\newcommand{\del}[0]{\partial}

\let\baraccent=\=
\renewcommand{\=}[1]{\stackrel{#1}{=}}

\newcommand{\Tr}[1]{\text{Tr}\left( #1 \right)}

\newcommand{\T}{{T^{1,1}}}
\newcommand{\ap}{\alpha'}

\newcommand{\abs}[1]{\left|#1\right|}
\newcommand{\calO}{\mathcal{O}}
\newcommand{\R}{\mathbb{R}}
\newcommand{\C}{\mathbb{C}}
\newcommand{\Z}{\mathbb{Z}}
\newcommand{\N}{\mathbb{N}}

\newcommand{\Pl}{\text{P}}
\newcommand{\IR}{\text{IR}}
\newcommand{\UV}{\text{UV}}

\newcommand{\CY}{\text{CY}}

\begin{document}
\thispagestyle{empty}

\rightline{DESY 18-212}

\vspace*{0.5cm}
\begin{center}
{\Large\bf  Thraxions: Ultralight Throat Axions}
\vspace{0.5cm}

{\large Arthur Hebecker,$\!^1$ Sascha Leonhardt,$\!^1$ Jakob Moritz,$\!^2$ Alexander Westphal$\,^2$\\[6mm]}

{\it
$^1\,$Institute for Theoretical Physics, University of Heidelberg, 
Philosophenweg 19,\\ 69120 Heidelberg, Germany\\[3mm]
$^2\,$Deutsches Elektronen-Synchrotron, DESY, Notkestra\ss e 85,\\
22607 Hamburg, Germany\\[3mm]
{\small\tt (\,a.hebecker, s.leonhardt~@thphys.uni-heidelberg.de\\
jakob.moritz, alexander.westphal~@desy.de\,)} }\\[.3cm]
\today
\end{center}

\vspace{1cm}

\begin{abstract}
We argue that a new type of extremely light axion is generically present in the type IIB part of the string theory landscape. Its mass is suppressed by the third power of the warp factor of a strongly warped region (Klebanov-Strassler throat), suggesting the name \textit{thraxion}. Our observation is based on the generic presence of several throats sharing the same 2-cycle. This cycle shrinks to zero volume at the end of each throat. It is hence trivial in homology and the corresponding $C_2$ axion is massive. However, the mass is warping-suppressed since, if one were to cut off the strongly warped regions, a proper 2-cycle would re-emerge. Since the kinetic term of the axion is dominated in the UV, an even stronger, quadratic mass suppression results. Moreover, if the axion is excited, the angular modes of the throats backreact. This gives our effective $C_2$ axion a finite monodromy and flattens its potential even further. Eventually, the mass turns out to scale as the third power of the warp factor. We briefly discuss possible implications for phenomenology and potential violations of the Weak Gravity Conjecture for axions. Moreover we identify a mechanism for generating super-Planckian axionic field ranges which we call \textit{drifting monodromies}. However, in the examples we consider, the potential oscillates on sub-Planckian distances in field space, preventing us from building a natural inflation model on the basis of this idea.
\end{abstract}

\newpage
\setcounter{tocdepth}{3} 
\tableofcontents

 \section{Introduction}
 
  Axion-like particles (axions for short) have become a main player in beyond-the-standard-model (BSM) physics in general \cite{Sikivie:2006ni,Jaeckel:2010ni} and in string phenomenology in particular \cite{Conlon:2006tq,Svrcek:2006yi,Grimm:2007hs,Arvanitaki:2009fg,Cicoli:2012sz}. They are relevant e.g.~in the context of the strong CP problem, inflation and as dark matter candidates. Furthermore, in the recent debate surrounding the Landscape/Swampland program \cite{Vafa:2005ui,Ooguri:2006in} and the Weak Gravity Conjecture (WGC) \cite{ArkaniHamed:2006dz}, axions have occupied a prominent place \cite{delaFuente:2014aca, Rudelius:2015xta, Montero:2015ofa, Brown:2015iha,Bachlechner:2015qja, Hebecker:2015rya, Heidenreich:2015wga, Junghans:2015hba, Kooner:2015rza}. This is also related to the question whether or with which parameters axion monodromy \cite{Silverstein:2008sg,McAllister:2008hb} can be realized in consistent quantum gravity settings \cite{Marchesano:2014mla, Blumenhagen:2014gta, Hebecker:2014eua, Ibanez:2015fcv,  Hebecker:2015zss, Klaewer:2016kiy}. 

  In this paper, we present a novel type of ultralight axion which, as we argue, is generically present in the type IIB part of the landscape, building on a proposal made in \cite{Hebecker:2015tzo}. Its extreme lightness, both in absolute terms and in relation to its decay constant (i.e.~compared to the scale $M_\Pl^4\exp(-M_\Pl/f)$ of generic non-perturbative potentials) lets it stand out among the many other stringy axions. It is surprising that, to the best of our knowledge, this very generic axionic degree of freedom was missed at the time when the string axiverse was being intensely studied. 

  Before turning to the details, we want to explain our central and, in our opinion, rather surprising, parametric results: Consider type IIB Calabi-Yau orientifold or F-theory models stabilized by fluxes and non-perturbative effects \cite{Giddings:2001yu, Kachru:2003aw, Balasubramanian:2005zx}. It is generally accepted that Klebanov-Strassler (KS) throats \cite{Klebanov:2000hb} with warp factor $w_\IR\ll 1$ will be present in an order-one fraction of such models \cite{Ashok:2003gk, Denef:2004ze, Hebecker:2006bn}. This warp factor can easily be {\it exponentially} small, such that it is justified to focus for the moment only on the dependence on $w_\IR$. In other words, let us for now set $R_{\CY}\sim M_\text{string}^{-1}$ and $N_\text{flux}\sim{\cal O}(1)$, thus ignoring all parameters except for the warp factor. Naively, the lightest states are then the glueballs (or warped-throat KK modes) with mass $\sim w_\IR$ (in Planck units). By contrast, we claim that an ultralight axion with mass $\sim w_\IR^3$ is frequently present in such settings. To be more precise, this happens at least in all cases where the fluxes stabilize the complex structure moduli near a conifold transition point in moduli space.
  
  Moreover, our axion has a decay constant $f\sim {\cal O}(1)$ in the simplest models\footnote{Note that despite the fact that strongly warped throats are needed to generate a small scalar potential for the axion, the decay constant is not suppressed by warping effects. This is because its internal field-profile is \textit{not} localized at the bottom of the throats, in contrast to some examples that have appeared in the literature~\cite{Dasgupta:2008hb,Franco:2014hsa}.}, which can be enhanced by products of flux numbers to super-Planckian values in more general settings, and an effective potential which can be much smaller than the naive expectation $V\sim \exp(-1/f)\,\cos(\phi/f)$ (again in Planck units). Clearly, this has potentially many interesting applications, from the WGC for axions to inflation and uplifting.
  
\begin{figure}[ht]
   \centering
   \def\svgwidth{0.65\textwidth}
   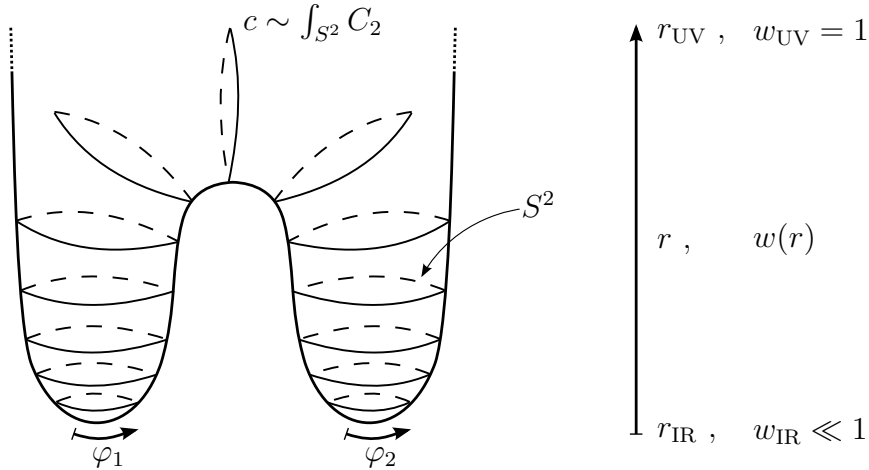
   \caption{An illustration of the setup of the double throat including the phases $\varphi_i$ and the axion $c$. The phases $\varphi_i$ describe physical rotations of each throat. We have not drawn the $S^3$ over every point of the double throat.}
   \label{fig:Intro_Setup}
  \end{figure}

  The paper is organized as follows. We start with the background solution in Sect.~\ref{sec:BackgroundSolution}. We consider a Calabi-Yau with a conifold point in complex structure moduli space at which multiple three-cycles degenerate simultaneously. We explain why this is a generic feature of Calabi-Yaus. Concentrating on the case of two degenerate three-cycles, we introduce separate deformation parameters $z_i$ with phases $\varphi_i = \arg{z_i}\,$, $i=1,2$, for the two deformed conifold regions. Crucially, the two conifolds, specifically the $S^3$-cycles describing the apices, are related in homology. As a result, the Calabi-Yau condition ensures that only one complex structure modulus, $z = z_1 = z_2$, is present. Deformations with $z_1 \neq z_2$ are massive. We then introduce fluxes stabilizing the complex structure modulus $z$ near the conifold point $|z| \ll 1$. The resulting geometry is illustrated in Fig.~\ref{fig:Intro_Setup}. One can see that the so-called $\mathcal{B}$-cycle is an $S^3$ which can be thought of as a family of $S^2$'s. This $S^2$ family reaches into both throats such that the $S^2$'s collapse at the apices. The corresponding dual $\mathcal{A}$-cycle is an $S^3$ over every point of the double throat in Fig.~\ref{fig:Intro_Setup}.
  
  In Sect.~\ref{sec:LocalBackReaction}, we introduce the axion\footnote{This may seem like a misnomer since the shift symmetry is completely broken: The 10d gauge invariance $C_2 \sim C_2 + \diff \Lambda_1$ does not imply a 4d shift symmetry because the two-sphere is trivial in homology $S^2 = \partial \Sigma_3$, $\int_{S^2} \diff \Lambda_1 = \int_{\Sigma_3} \diff^2 \Lambda_1 = 0$. However, from the unwarped UV-perspective the two-sphere is non-trivial and the field is a proper axion, with a monodromy created only by the warped-down IR region.} $c \sim \int_{S^2} C_2$, called \textit{thraxion} from now on. An excursion of the thraxion generates non-zero opposite values of the RR-field strength $F_3$ at the ends of the two throats. Local backreaction of the resulting energy density then deforms the two throats independently: While the phase $\varphi_1$ of the local deformation parameter of one throat is displaced by fluxes, the phase $\varphi_2$ of the other throat is displaced in the opposite direction by anti-fluxes. This breaks the constraint $\varphi_1 = \varphi_2$ coming from the CY condition and the homology relation between the two throats. In Sect.~\ref{sec:CYBreakingPotential}, we calculate the potential induced by non-vanishing 10d Ricci curvature that stabilizes the two deformation parameters against each other. After integrating out heavy degrees of freedom, the result is an effective potential for the thraxion with the properties described above and discussed in Sect.~\ref{sec:FirstDiscussion}.
  
  Sect.~\ref{sec:4dSugra} rederives the effective axion potential from a proposed generalization of the Gukov-Vafa-Witten (GVW) superpotential \cite{Gukov:1999ya} that includes the axion. We thereby reproduce the results in 4d supergravity language, and identify the \textit{saxion} partner of the axion. In Sect.~\ref{sec:Multithroat} we generalize our results to general multi throat systems where one or more ultra-light thraxions appear.
  
  We discuss the consistency of our results with the holographic dictionary in the KS context \cite{Klebanov:2000nc,Klebanov:2000hb} in Sect.~\ref{sec:gaugegravity} by matching the enhancement of the decay constant of our axion $\int C_2$ with gaugino condensation on the gauge theory side. Applications and implications of these results are the content of Sect.~\ref{sec:Applications}. We consider as an explicit example the quintic three-fold stabilized near a conifold transition point. We study the scalar potential for judicious choices of flux quanta. Interestingly, the overall monodromy enhancement is given by the least common multiple of all the flux quanta which can easily become parametrically super-Planckian. However, the presence of sub-Planckian modulations generically prevents successful slow-roll inflation. The underlying idea is that a large monodromy is generated by unsynchronized phases (of monodromies of individual throats) drifting away from one another. We call this mechanism \textit{drifting monodromies}. This mechanism can also be thought of as the well-known beat phenomenon in accoustics, in which the interference of harmonics with slightly different small wavelengths leads to large wavelength oscillations. For related alternative possibilities of generating large decay constants see \cite{Kim:2004rp,Berg:2009tg,Hebecker:2015rya,Palti:2015xra,Choi:2015fiu,Kaplan:2015fuy,Blumenhagen:2016bfp,Buratti:2018xjt,Hebecker:2018a,Hebecker:2018fln}\footnote{Note in particular the following two papers: The work of \cite{Blumenhagen:2016bfp} is closely related to ours in making use of the conifold complex structure modulus $z$ to create super-Planckian decay constants, while on the technical level the approach is very different. Ref.~\cite{Buratti:2018xjt} defines the 5d axion $\int B_2$ on the KT background, analogously to our thraxion. There, the geometric backreaction via the 5d breathing mode allows for monodromy-induced super-Planckian field ranges to be explored in an anisotropic and inhomogeneous 5d spacetime.}. We also describe a clash with the WGC: The effective Euclidean instanton action determined from the scale of the effective potential violates the axionic WGC ($S \lesssim q M_\Pl/ f_\text{eff}$) parametrically, but instead a slightly weaker inequality holds ($S\lesssim (q M_\Pl/f_\text{eff})^2$). After commenting on the relevance of our findings to light axion phenomenology, we finally consider interesting possibilities for uplifting to de Sitter vacua. We draw our conclusions in Sect.~\ref{sec:Conclusion}.

 \section{Backreacted Potential of the Thraxion from 10d}
 \label{sec:Upshot}
 
 \subsection{Geometric and Flux-Background}
 \label{sec:BackgroundSolution}
 
 \subsubsection{Geometric Features of Generic CYs}
 \label{sec:GeometricBackground}
  
  First we will explain the basic geometric requirements for our discussion to apply. We will explain why we expect them to be \textit{generically} met.
  
  Let us consider compactifications of type IIB string theory on a Calabi-Yau (CY) threefold, which leads to an effective $\mathcal{N}=2$ supergravity theory in four dimensions. There is a moduli space of vacua parametrized by the $h^{2,1}$ complex structure moduli and $h^{1,1}$ K\"ahler moduli. There are special points in complex structure moduli space called conifold points where the CY develops conical singularities, and one or more three-cycles degenerate to zero volume \cite{Candelas:1989ug,Candelas:1989js}. We will consider a CY near such a conifold point, where \textit{multiple} three-cycles degenerate.
  
  To understand in what sense this is generic, we consider cases in which it is also possible to \textit{resolve} conifold singularities while preserving the CY condition. This does not restore the degenerate three-cycles to finite size, but rather produces non-trivial two-cycles. In going through this so-called \textit{conifold transition}, a new CY threefold with Hodge numbers $\tilde{h}^{1,1}>h^{1,1}$ and $\tilde{h}^{2,1}<h^{2,1}$ is produced \cite{Candelas:1988di}. Whenever two CY threefolds are connected via a conifold transition, at the conifold transition point two or more three-cycles $\mathcal{A}_i$ that are related in homology shrink to zero size. This is a consequence of demanding K\"ahlerity on the resolved side of the transition \cite{Candelas:1990rm}. It is widely believed that a \textit{generic} CY threefold is related to other CY threefolds via conifold transitions \cite{Reid:1987,Candelas:1988di}. While research on this is still on-going, there are large classes of CY's for which this has been shown \cite{Chiang:1995hi,Avram:1997rs,Batyrev:1994hm,Batyrev:2008rp,Lynker:1995sy,Batyrev:1998kx}\footnote{Note that this does not mean that every conifold singularity (or even a generic one) is also such a transition point. For example, the mirror quintic threefold at vanishing complex structure has a single shrunken three-cycle. Hence there is no resolved CY geometry \cite{Candelas:1990rm}.}. Therefore, a generic CY has loci in complex structure moduli space where multiple three-cycles $\mathcal{A}_i$ degenerate together. We expand on this in Sect.~\ref{sec:4dSugra}. Being related in homology, the number of homology classes is smaller than the number of collapsing three-cycles. For now we focus on the case of precisely two cycles $\mathcal{A}_{1,2}$ that degenerate. From the above it immediately follows that they are related in homology $[\mathcal{A}] \equiv [\mathcal{A}_{1}] = [\mathcal{A}_{2}]$. There is a symplectic dual three-cycle $\mathcal{B}$ connecting the two singular points. We will call this system a \textit{double conifold}. Its complex structure will be denoted by $z$ and the double conifold singularity develops in the limit $|z|\to 0$.
  
\begin{figure}[ht]
   \centering
   \def\svgwidth{0.7\textwidth}
   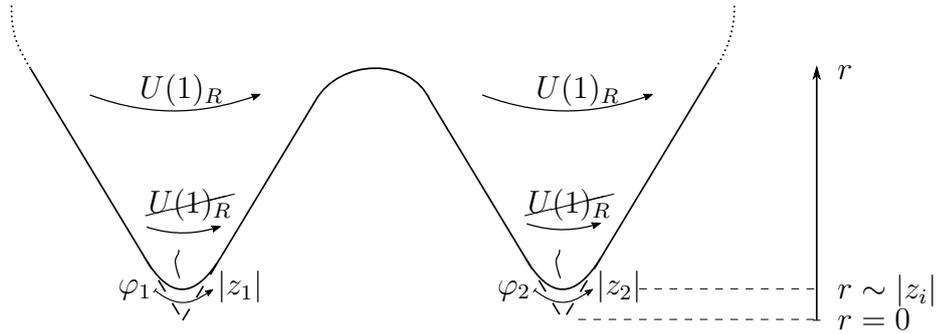
   \caption{The double conifold with asymptotic $U(1)_R$ symmetric regions.}
   \label{fig:Upshot_DoubleConifold}
  \end{figure}

  We introduce the fields $z_1$ and $z_2$ as illustrated in Fig.~\ref{fig:Intro_Setup}. These fields may be thought of as `local complex structure deformations' $z_i = \int_{\mathcal{A}_i} \Omega$, with the holomorphic three-form $\Omega$ of the CY, and describe independent \textit{local} deformations of the manifold near one of the two apices. Thus, in the vicinity of either apex of the double conifold we want to describe the manifold by embedding it into $\C^4$ via
  \begin{equation}
   w_1^2 + w_2^2 + w_3^2 + w_4^2 = z_i \,, \quad w \in \C^4 \,.
   \label{eq:DeformedConifold}
  \end{equation}
  While the homology relation $[\mathcal{A}_1]=[\mathcal{A}_2]$ enforces $z = z_1 = z_2$ on complex structure moduli space\footnote{This is because the difference $\mathcal{A}_1-\mathcal{A}_2$ is the boundary of a $4$-chain $\mathcal{C}$. Therefore, one has $z_1-z_2=\int_{\mathcal{A}_1}\Omega-\int_{\mathcal{A}_2}\Omega=\int_{\del\mathcal{C}}\Omega=\int_{\mathcal{C}}\diff\Omega$. On complex structure moduli space one has $\diff\Omega=0$, and hence $z_1=z_2$.}, we will also consider deformations of the manifold such that $z_1 \neq z_2$, i.e.~deformations away from complex structure moduli space\footnote{For similar considerations with deformations away from K\"ahler moduli space, i.e.~$\diff J\neq 0$, see ref. \cite{Grimm:2008ed}.} (i.e.~$\diff\Omega\neq 0$). It is important to note that a deformation of the local phases $\varphi_i$ of $z_i$ becomes an isometry far away from the tip of either conifold: The cross-section of either deformed conifold at a given radial coordinate $r$ possesses a spontaneously broken $U(1)_R$ isometry\footnote{The index $R$ is due to this symmetry group being the R-symmetry of the dual supersymmetric gauge theory. While this is of no importance to us, we keep it for notational consistency.} realized as a rotation of the local deformation parameter $z_i$ \cite{Klebanov:1998hh}, see Fig.~\ref{fig:Upshot_DoubleConifold}. In the limit of large radial coordinates, $r^3/\abs{z_i} \to \infty$, the deformed conifold becomes indistinguishable from the singular conifold. Therefore, the symmetry remains unbroken in this limit. We make this explicit in App.~\ref{sec:Notation}.
  
\begin{figure}[ht]
   \centering
   \def\svgwidth{0.5\textwidth}
   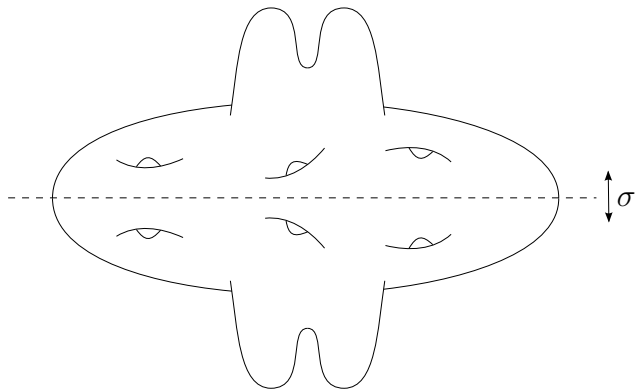
   \caption{A sketch of the the orientifold projection $\sigma$. It maps the two originally independent double throat cycles $\mathcal{B}$ and $\mathcal{B}'$ onto one another.}
   \label{fig:Orientifold}
  \end{figure}

  To complete the discussion of the geometric setting, we note that one may go beyond the simplest case with exactly two collapsing three-cycles. Such multi conifold situations are analyzed in Sect.~\ref{sec:4dSugra}. Furthermore, for reasons of tadpole cancellation we are interested in CY threefolds which are orientifolded such that O3/O7-planes arise. This projection should leave the conifold transition intact and preserve the key ingredient of a ${\cal B}$-cycle reaching down into several conifold regions. In the double throat case, this is realized if two originally present pairs of throats are mapped to each other by the orientifold projection, see Fig.~\ref{fig:Orientifold}. This is completely analogous to the widely-discussed double throat system of the oldest axion-monodromy models, see e.g.~\cite{McAllister:2008hb,Retolaza:2015sta} (just simpler, since we need no 2-cycle for the NS5 brane and can hence use standard KS throats). More generally, F-theory solutions with the analogous geometric properties can be considered. Here the tadpole cancellation relies on the fourfold Euler number and no orientifolding is required. Either way, we do not expect that the orientifolding condition or the fourfold embedding endangers the generality of our setting.

 \subsubsection{Moduli Stabilization by Fluxes}
  
  As was shown in \cite{Giddings:2001yu}, generic choices of three-form flux quanta stabilize the axio-dilaton as well as all the complex structure moduli. Whenever the flux quanta $K\equiv -\frac{1}{(2\pi)^2 \alpha'}\int_{\mathcal{B}}H_3$ and $M\equiv \frac{1}{(2\pi)^2 \alpha'} \int_{\mathcal{A}_1}F_3=\frac{1}{(2\pi)^2 \alpha'}\int_{\mathcal{A}_2}F_3$ satisfy\footnote{It has recently been claimed that achieving this hierarchy may be impossible \cite{Bena:2018fqc}. We think that the assumptions of \cite{Bena:2018fqc} are too restrictive (as is pointed out by the authors themselves). In particular, O7-planes with curvature contribute to the D3 tadpole and huge Euler numbers are achievable in F-theory. We expect that this will allow to make the tadpole large enough for our purposes.} $K\gg g_s M$ (where $g_s$ is the string coupling), our complex structure modulus $z = \abs{z} e^{i \varphi}$ is stabilized near the conifold point
  \begin{equation}
   |z|\propto \exp\left(-2\pi \frac{K}{g_s M}\right)\ll 1\,.
  \end{equation}
  Separating the equation $D_z W = 0$ into real and imaginary parts (see Sect.~3 of \cite{Giddings:2001yu}), we also find that the phase is stabilized. Its value is set by the RR-3-form flux $Q = \frac{1}{(2\pi)^2 \alpha'}\int_{\mathcal{B}}F_3$
  \begin{equation}
   \varphi = 2 \pi \frac{Q}{M}\, .
   \label{eq:FluxStabilizedValues}
  \end{equation}
  Locally, we can always set $\varphi$ to $0$ by an appropriate redefinition of the angle. Conversely, without loss of generality, we will choose $Q=0$.
  
  Moreover, backreaction of fluxes leads to the formation of warped throats (or Klebanov-Strassler throats). Within these, the metric is well approximated by the Klebanov-Tseytlin solution\footnote{Near the bottom of the throat, it has to be replaced by the full Klebanov-Strassler solution \cite{Klebanov:2000hb}.} \cite{Klebanov:2000nc}
  \begin{equation}
  \diff s^2=w(r)^2 \eta_{\mu \nu} \diff x^\mu \diff x^\nu+w(r)^{-2}(\diff r^2+r^2 \diff s^2_{T^{1,1}})\, , \quad w(r)^2\sim \frac{r^2}{g_sM\alpha'}\log(r/r_{\IR})^{-\frac{1}{2}}\, ,
  \label{eq:KTMetric}
  \end{equation}
  with radial coordinate $r$ and warp factor $w(r)$. The radial coordinate is cut off in the IR by the Klebanov-Strassler region and in the UV by the gluing into the bulk CY. One has $w_{\IR}\equiv w(r_{\IR})\sim r_{\IR}/r_{\UV}\sim |z|^{1/3}$ giving rise to an exponential hierarchy à la Randall-Sundrum \cite{Randall:1999ee,Klebanov:2000hb,Giddings:2001yu}. As we have explained, in the vicinity of a conifold transition point a \textit{double throat} (or even \textit{multi throat}) forms\footnote{It may not seem obvious that the units of NS-flux on the $\mathcal{B}$-cycle are split democratically so that \textit{each} conifold region is replaced by a warped throat. In fact we will see that there is a light dynamical field that controls this relative distribution (see Sect.~\ref{sec:4dSugra}). In the vacuum however this field is stabilized such that fluxes are indeed distributed democratically.}, see Fig.~\ref{fig:Intro_Setup}.
  
  The three-cycle $\mathcal{B}$ can be thought of as an $S^2$ fibered over the radial direction of the conifold \cite{Candelas:1989js}. The $S^2$ collapses at the two \textit{tips} of the deformed conifolds. As discussed in \cite{Hebecker:2015tzo}, there exists a 4d mode $c(x)$ on the double throat background that can be thought of as the integral of the Ramond-Ramond (RR) two form $C_2$ over the $S^2$ as measured far away from the tips of the double throat. A non-trivial field excursion leads to the creation of regions with flux on the two respective ends of the cycle $\mathcal{B}$, and hence a red-shifted potential $V(c)=\frac{1}{2}m^2c^2+...$, with $m^2\sim w_{\IR}^4$. Backreaction of the geometry was neglected in \cite{Hebecker:2015tzo}.
  
\begin{figure}[ht]
   \centering
   \def\svgwidth{0.7\textwidth}
   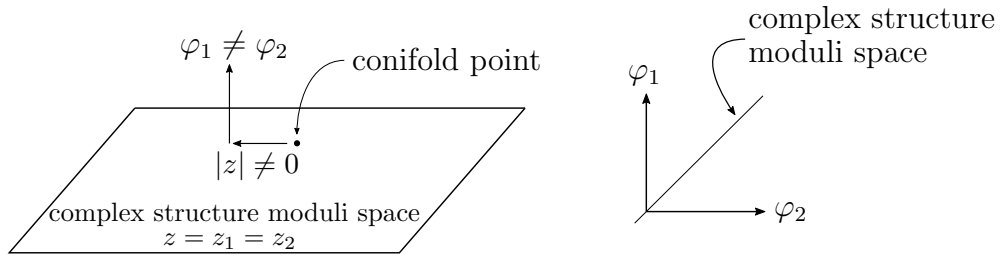
   \caption{Illustration of the $z_1$-$z_2$ deformation space. The complex structure moduli space is the subspace $z_1 = z_2$. We only consider deformations away from $z_1 = z_2$ outside the conifold point.}
   \label{fig:Upshot_DeformationSpace}
  \end{figure}

  In the rest of this section, we will establish the following points:
  \begin{itemize}
   \item The fields $z_1$ and $z_2$ of the two respective throats adjust to the flux/anti-flux pair in such a way that within the two throats supersymmetry is restored locally.
   \item This adjustment of $z_1$ and $z_2$ takes us away from the complex structure moduli space, which is characterized by $z_1 = z_2$ (cf. Fig.~\ref{fig:Upshot_DeformationSpace}). For $z_1 \neq z_2$, the CY condition is broken and a scalar potential is generated. This potential is of the order $|z|^2\sim w_{\IR}^6$ and receives its dominant contributions from the bulk CY.
   \item The backreacted scalar potential is periodic in $c$ with periodicity $2\pi M$. Hence, the naive $2\pi$ periodicity of the $c$-axion is enhanced by a \textit{finite} factor $M$. While this does \textit{not} allow for a super-Planckian effective axion decay constant $f\gg M_\Pl$, approximately Planckian values are possible (see however Sect.~\ref{sec:Applications} for a way to generate large axion periodicities).
  \end{itemize}
 
 \subsection{Local Backreaction in the Throat}
 \label{sec:LocalBackReaction}
  
\begin{figure}[ht]
   \centering
   \def\svgwidth{0.5\textwidth}
   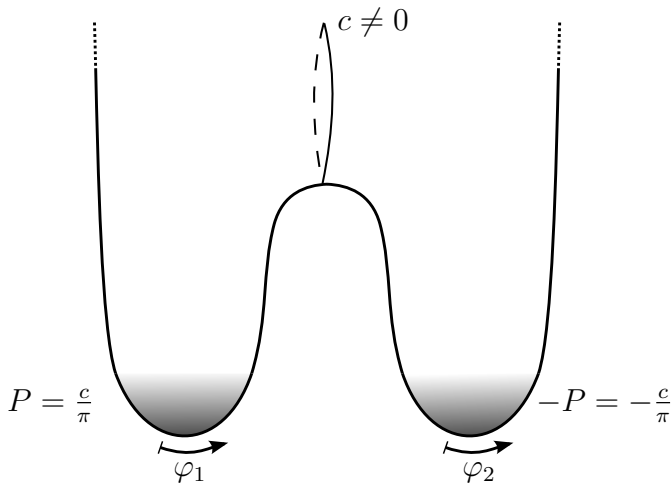
   \caption{Fluxes induced by non-vanishing $c$ are localized at the tips of the throats.}
   \label{fig:Upshot_Fluxes}
  \end{figure}

  We start by discussing how a single throat reacts locally to a finite field excursion $c$. Since the outcome will be that the throat almost perfectly adjusts to produce a locally supersymmetric configuration, we are entitled to use the 4d description in terms of the GVW superpotential \cite{Gukov:1999ya,Giddings:2001yu} for two KS throats. As far as (say) the first local throat is concerned, a non-vanishing field excursion $c$ cannot be distinguished from additional flux $P\equiv c/2\pi$ on the local portion of the $\mathcal{B}$-cycle\footnote{We will use the term flux also for the non-quantized integral $\int F_3$ over some region.}, see Fig.~\ref{fig:Upshot_Fluxes},
  \begin{equation}
   c=\frac{1}{2\pi \alpha'}\int_{S^2}C_2=\frac{1}{2\pi \alpha'}\int_{\frac{1}{2}\mathcal{B}}\diff C_2=(2\pi)P\, .
   \label{eq:LocalFlux}
  \end{equation}
  Considering a single throat with complex structure modulus $z_1\equiv |z_1|e^{i\varphi_1}$, the arguments of GKP \cite{Giddings:2001yu} show that there are SUSY configurations for 
  \begin{equation}
   \varphi_1 = 2 \pi \frac{P}{M} =c/M\, ,
   \label{eq:SUSYvac}
  \end{equation}
  compare \eqref{eq:FluxStabilizedValues}. Hence, the throat can locally relax the SUSY breaking induced by the extra RR-flux by adjusting the phase of the deformation parameter $z_1$. However, the second throat sees the field excursion $c$ as the flux $-P = -c/2\pi$ on the $\mathcal{B}$-cycle for which there exists a locally supersymmetric configuration with $\varphi_2=-c/M$.
  
  Since there is no additional flux that would lead to $\abs{z_1} \neq \abs{z_2}$, we keep $\abs{z} = \abs{z_1} = \abs{z_2}$ fixed at the stabilized value \eqref{eq:FluxStabilizedValues} for what follows. These two modes decouple from the discussion at hand.
  
  We can encode the discussion above in a 4d EFT potential. To quadratic order, the discrepancy between the local fluxes and local deformations induces a potential
  \begin{equation}\label{eq:fluxpotential}
   V_\text{flux}(c,\varphi_1,\varphi_2)= \frac{1}{2}\mu^4 (M\varphi_1-c)^2+\frac{1}{2}\mu^4 (M\varphi_2+c)^2\, ,
  \end{equation}
  We have $\mu\sim w_{\IR}$ since the potential is generated locally near the tip of the throats.
  
  The fact that only the combinations $M\varphi_i\pm c$ appear in the scalar potential can be derived also via ten-dimensional considerations. As shown in App.~\ref{sec:AxionPeriodicity}, in the local throats, the combined transformation $\varphi_{1,2}\longrightarrow \varphi_{1,2}\pm \delta$, $c\longrightarrow c+M\delta$ is a diffeomorphism acting on the KS solution. Hence, only the invariant combinations $M\varphi_{1,2}\mp c$ can appear in the scalar potentials that are generated locally at the bottom of the throats.
  
  The potential derived so far possesses a flat direction which we parametrize by $c$. This flat direction is given by
  \begin{equation}
   \varphi_{1} = - \varphi_{2} =c/M \,.
   \label{eq:FieldsAtMinimum}
  \end{equation}
 
 \subsection{The CY Breaking Potential}
 \label{sec:CYBreakingPotential}
  
  In the preceding section we have argued that the individual throats react to the field excursion $c$ by adjusting their local deformation parameters $z_1$ and $z_2$, more specifically their phases $\varphi_1$ and $\varphi_2$ respectively. Since the corresponding CY has only one complex structure modulus $z \equiv  z_1 \equiv z_2$, the mode $z_1/z_2$ or rather $\varphi_1 - \varphi_2$ must be massive already before fluxes are turned on. This eliminates the remaining flat direction in the potential.
  
  We parametrize the part of the scalar potential that is due to the breaking of the CY condition as
  \begin{equation}
   V_\text{CY-breaking}=\Lambda^4 (1-\cos(\varphi_1-\varphi_2))\, ,
   \label{eq:CYBreakingPot}
  \end{equation}
  with a yet undetermined scale $\Lambda$. In writing this we have assumed the following:
  \begin{enumerate}
   \item The potential is a function of the difference $\varphi_1-\varphi_2$ only.
   \item It satisfies $V_\text{CY-breaking}(\varphi_1-\varphi_2) = V_\text{CY-breaking}(\varphi_1-\varphi_2 + 2 \pi)$.
   \item The lowest harmonic dominates.
  \end{enumerate}
  Condition a) must hold because only the local fluxes of the throats stabilize $\varphi_{1,2}$ individually and, without the flux potential, the complex structure modulus $\varphi = (\varphi_1 + \varphi_2)/2$ should be a flat direction. We expect condition b) to hold because we see no reason for a monodromy. Condition c) is a rather unimportant assumption that we make for ease of exposition.
  
  We combine \eqref{eq:fluxpotential} and \eqref{eq:CYBreakingPot} and integrate out $\varphi_{1,2}$ under the assumption $\Lambda^4 \ll \mu^4$ (to be justified below). This corresponds to imposing \eqref{eq:FieldsAtMinimum}. The effective potential takes the form
  \begin{equation}
   V_\text{eff}(c)=\Lambda^4\left(1-\cos(2c/M)+\mathcal{O}(\Lambda^4/\mu^4)\right) \, .
  \label{eq:vc}
  \end{equation}
    
  The height of this potential can be estimated using the 10d solution. To do so, we need to develop a clear picture of how field profiles and 10d geometry change if we excite $c$. Recall that $c$ is originally defined by a particular `Wilson line' VEV of $C_2$ in the UV of the two throats (as well as in the piece of the CY connecting them). Turning on this VEV and focussing on one throat only, we observe a backreaction of the throat geometry which maintains SUSY and corresponds to the motion along a flat direction in 4d field space. This is independently true for the second throat, which backreacts in the opposite way: $\varphi_1=-\varphi_2=c/M$. 

\begin{figure}[ht]
   \centering
   \def\svgwidth{0.5\textwidth}
   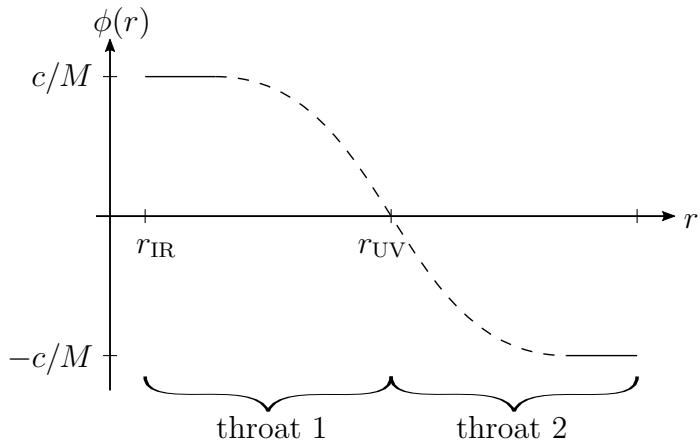
   \caption{The expected profile of the 10d/5d mode along the radial direction.}
   \label{fig:Upshot_Profile}
  \end{figure}

  Now, the crucial point is that these rotations, defined by the IR parameters $\varphi_{1,2}$, must by continuity be accompanied by a corresponding $r$-dependent rotational profile along the whole throat. We encode this in a five-dimensional field $\phi(x^\mu,r)$ that interpolates between $\varphi_1= c/M$ and $\varphi_2=-c/M$ at the respective ends of the throats. This is illustrated in Fig.~\ref{fig:Upshot_Profile}, which also displays the expected symmetry: The solution should be antisymmetric under the exchange of the two throats.\footnote{By a slight abuse of notation we stick with the familiar variable $r$, although according to our figure this variable must now be growing as one goes down the second throat.}
  
  For computational simplicity, we model the transition region between the throats by a single point, $r=r_\UV$\footnote{In fact, the exact UV geometry and UV fluxes are irrelevant as long as we do not consider perturbative and non-perturbative corrections \cite{Baumann:2010sx}.}. In doing so, we ignore effects of the unwarped CY region (accepting an ${\cal O}(1)$ error). The field $\phi$ must be zero at this point for symmetry reasons. This symmetry also ensures that we may limit our attention to one of the two throats when computing the energy density associated with an excursion of $c$.

  The key point is that, after these preliminaries, we are actually able to estimate this energy. It is given by the gradient energy of $\phi$, which accounts precisely for the clash between the opposite rotations of $\varphi_1$ and $\varphi_2$. The relevant action for $\phi=\phi(x^\mu,r)$ is obtained by dimensionally reducing the 10d Ricci scalar to quadratic order on the warped conifold background (see App.~\ref{sec:ExactResults}):
  \begin{equation}
  \begin{split}
   S[\phi] =& \frac{M_\text{10d}^8}{2} \int d^4 x \int_{r_{\IR}}^{r_{\UV}} \diff r ~ \sqrt{-g_\text{5d}} \, r^5 w(r)^{-5} ~ \epsilon(r)^2 ~ \left(-\frac{1}{2} g_\text{5d}^{MN} \partial_M \phi \partial_N \phi\right) \\
   =& \frac{M_\text{10d}^8}{2} \abs{z}^2 \int d^4 x \int_{r_{\IR}}^{r_{\UV}} \frac{\diff r}{r} \left(-\frac{1}{2}(\del_r \phi)^2 - \frac{1}{2} w(r)^{-4}(\del_\mu \phi)^2\right)\, .
   \label{eq:5dActionForPhi}
  \end{split}
  \end{equation}
  Considering the metric \eqref{eq:KTMetric}, this form of the 5d action is easily understood. The metric naturally splits into a 5d part $g_\text{5d}$ in the external and radial direction and an angular part $\propto g_{\T}$. The latter contributes the $r$-dependent terms $\sqrt{g_{\T}} \propto r^5 w(r)^{-5}$ to the metric determinant. The function $\epsilon(r)$ encodes the degree of $U(1)_R$ symmetry breaking, compare Fig.~\ref{fig:Upshot_DoubleConifold}. Since a field excursion $\phi$ is obtained by acting with a $U(1)_R$ transformation, any terms in the action that contain the field must be multiplied by the factor $\epsilon(r)^2$. This symmetry breaking is due to the deformation near the tip and takes the form $\epsilon(r) \sim  \abs{z}/r^3$ far from the tip of the throat, see App.~\ref{sec:Notation}.

  We now apply the static approximation (i.e.~disregard the $(\partial_\mu\phi)^2$ in (\ref{eq:5dActionForPhi})), derive the equation of motion and solve it for the boundary conditions $\phi(x^\mu,r_{\rm UV})=0$ and $~\phi(x^\mu,r_{\rm IR})=\varphi_1=c/M$. This gives 
  \begin{equation}
   \phi(x^\mu,r)= c/M \frac{r_{\UV}^2-r^2}{r_{\UV}^2-r_{\IR}^2}\label{phis}
  \end{equation}
  which, inserting back in (\ref{eq:5dActionForPhi}), leads to a 4d potential $V \sim \abs{z}^2 c^2$. This is a result at quadratic order in $c$ but, comparing to \eqref{eq:vc}, this is sufficient to infer that $\Lambda^4 \sim \abs{z}^2$. Finally inserting the stabilized value $\abs{z} \propto w_{\IR}^3$ \cite{Giddings:2001yu}, we arrive at $\Lambda^4 \sim w_\IR^6$. Our assumption $\mu^4\gg \Lambda^4$ is now \textit{a posteriori} justified. It is also apparent that the effective mass of our ultra-light field is $m_c\sim w_{\IR}^3$. 

  Let us justify the use of the static approximation above. Really, we should have supplemented (\ref{eq:5dActionForPhi}) by the kinetic term $S_\text{kin}[c] \sim \int \diff^4x\,(\partial_\mu c)^2$, imposed the constraint $\phi(x^\mu,r_{\rm UV})=c(x^\mu)/M$, and determined the mass of the lowest-lying KK mode of the resulting 5d action. However, it is intuitively clear that the UV-dominated kinetic term of $c$ is much more important than the warped-down 4d-gradient term $(\partial_\mu\phi)^2$ in (\ref{eq:5dActionForPhi}). Thus, $c$ is the most inert part of the system and it is an excellent approximation to assume that the $\phi$ profile extremizes just the 5d-gradient-part of the action. To make this quantiative, one may substitute $c$ on the r.h. side of (\ref{phis}) with the plane wave $c=\exp(ikx)$ (with $k^2=-m_c^2$) and check that the resulting $(\partial_\mu\phi)^2$ contribution from (\ref{eq:5dActionForPhi}) is negligible compared to $S_\text{kin}[c]$.\footnote{Note that there is no contribution from $S_\text{pot}[c] \sim \int \diff^4x \diff r \,(\partial_r c)^2$. When exciting $c$ in the UV, the local deformation parameter adjusts as explained above. The profile $c(r)$ is now stabilized in turn, for energies $\Lambda < E < \mu$, to the radially constant value of the phase of the deformation parameter $\pm M \varphi_i$. We expand on this in App.~\ref{sec:AxionPeriodicity}.}
  
  More details are given in App.~\ref{sec:ExactResults}.
  
 \subsection{Discussion of Results}
 \label{sec:FirstDiscussion}
  
  The information we have gathered can be summarized in an effective Lagrangian\footnote{For ease of exposition we have written down a diagonal kinetic matrix. This is not quite the case but is not relevant for our discussion. See App.~\ref{sec:AxionPeriodicity} for details.},
  \begin{equation}
  \begin{split}\label{eq:eff.Lagr1}
   \mathcal{L}=&-\frac{1}{2}f_{\varphi}^2 (\del \varphi_1)^2-\frac{1}{2}f_{\varphi}^2 (\del \varphi_2)^2-\frac{1}{2}f_c^2(\del c)^2 \\
   &-\frac{1}{2}\mu^4 (M\varphi_1-c)^2-\frac{1}{2}\mu^4 (M\varphi_2+ c)^2-\Lambda^4(1-\cos(\varphi_1-\varphi_2))\, ,
  \end{split}
  \end{equation}
  with coefficients
  \begin{equation} 
  \begin{aligned}
  	f_\varphi^2 ~ \sim & ~\log(w_\IR^{-1})^{-3/2}w_\IR^2 ~ M_\Pl^2\, ,\quad 
  	&&f_c^2 ~\approx ~ \frac{2}{9M^2}\log(w_\IR^{-1})^{-1} ~ M_\Pl^2\,,\\
  	\Lambda^4 ~\sim &~ \frac{g_s^2}{(g_sM)^4}\log(w_\IR^{-1})^{-7/2}w_\IR^6 M_\Pl^4\, ,\quad 
  	&&\mu^4 ~\sim ~ \frac{g_s^4}{(g_sM)^{6}}\log(w_\IR^{-1})^{-3}w_\IR^4 ~ M_\Pl^4\, .
  	\label{eq:coefficients}
  \end{aligned}
  \end{equation}
  The above expressions are valid for the special case of the bulk CY having a single characteristic length-scale, $R_\CY^6 \sim \text{Vol}(\CY)$, and where the throats marginally fit into the bulk, i.e.~$R_{\CY}^4\sim R_{\text{throat}}^4\sim g_sMK \alpha'^2$. For the general case see App.~\ref{Appendix:AxionDecayConstant}. We briefly pause to explain what is meant by the requirement that the throats fit into the bulk marginally: On the one hand the bulk CY has an overall size $R_{\CY}$ that is set by a combination of the K\"ahler moduli. On the other hand the throats can really be thought of as objects of a characteristic physical size $R_{\text{throat}}$ embedded into the bulk CY. This size $R_{\text{throat}}^4$ is well known to be set by the local D$3$ brane charge stored in the fluxes of the throat \cite{Klebanov:2000hb,Giddings:2001yu}, independently of the size of the bulk CY. For this to be a geometrically consistent configuration, we should require $R_{\CY}> R_{\text{throat}}$. Taking $R_{\CY}\sim R_{\text{throat}}$ is the case where the throats fit into the bulk CY only \textit{marginally}.\footnote{For a recent discussion on this see \cite{Carta:2019rhx}.}
  
  Far below the scale $w_{\IR}M_\Pl$ we may integrate out $\varphi_{1,2}$, to obtain the effective Lagrangian
  \begin{equation}\label{eq:eff.Lagr2}
  \mathcal{L}'=-\frac{1}{2}f_c^2 (\del c)^2-\Lambda^4 (1-\cos(2c/M))\, .
  \end{equation}
  We would like to highlight the following points,
  \begin{itemize}
   \item Our simplification $R_\text{throat} = R_{S^2} = R_\CY$ gives the largest possible value for the decay constant $f_\text{eff} = M f_c$; any hierarchy $R_\text{throat} < R_{S^2} < R_\CY$ suppresses its value. Taking into account logarithmic corrections, the maximal periodicity one can achieve is $\mathcal{O}(M_\Pl/\sqrt{\log w_{\IR}^{-1}})$ (see App.~\ref{Appendix:AxionDecayConstant}). A large hierarchy $w_{\IR}\ll 1$ suppresses the periodicity only very mildly. By taking $g_s M$ and $g_s^{-1}$ to be large, the 10d perturbative expansion becomes better controlled without affecting the axion periodicity. In this sense, our axion can be made approximately Planckian.
   \item The mass of the axion is $\mathcal{O}(w_{\IR}^3)$ which is parametrically smaller than both the warped Kaluza-Klein scale ($\mathcal{O}(w_{\IR})$), and the estimate of \cite{Hebecker:2015tzo}, where backreaction of the local geometry was not taken into account ($\mathcal{O}(w_{\IR}^2)$). The mass-spectrum is essentially \textit{gapped}.
   \item As pointed out before, the scale of the effective potential is set by the $U(1)_R$ breaking induced by the deformation of the conifold as measured in the UV, $\epsilon^2(r_{\UV}) \propto \abs{z}^2$. Strictly speaking this is \textit{not} a warp factor suppression, although for moderate CY volumes $|z|$ and $w_{\IR}^3$ are of the same order\footnote{One might for instance be tempted to consider the large volume limit where warping becomes negligible. In this case the scale of the potential would still be given by $|z|^2\ll 1$.}.
  \end{itemize}
  The following caveats should be noted: The effective Lagrangians \eqref{eq:eff.Lagr1} and \eqref{eq:eff.Lagr2} are incomplete: We have worked in the regime of classical type IIB solutions so at least the universal K\"ahler modulus $T$ is not yet stabilized. Moreover, we have not included the $b$-axion that complexifies $c$. Finally, there is no parametric separation between the mass scale of the complex structures and the warped Kaluza-Klein scale. Hence, the Lagrangian \eqref{eq:eff.Lagr1} does not define a useful effective field theory in the Wilsonian sense. Equation \eqref{eq:eff.Lagr2} however \textit{does} give rise to a Wilsonian effective Lagrangian once it is completed by the $b$-axion and the K\"ahler modulus $T$.

 \subsection{The \texorpdfstring{$B_2$}{B2}-axion}\label{sec:b-axion0}
 
\begin{figure}[ht]
  	\centering
  	\def\svgwidth{0.7\textwidth}
  	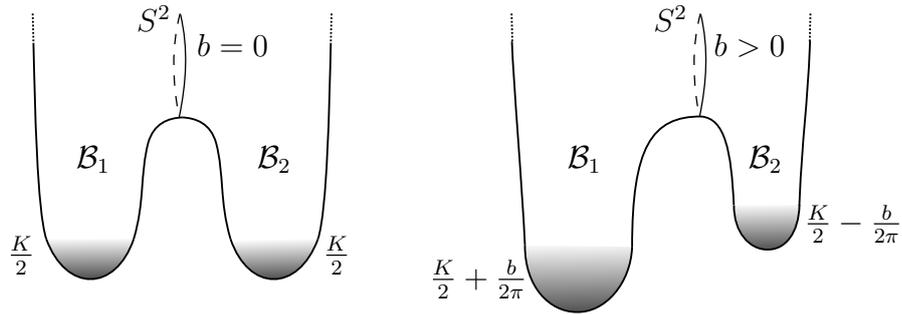
  	\caption{The physical effect of a field excursion of the $b$-axion in the double throat system. One throat becomes shorter, whereas the other becomes longer.}
  	\label{fig:BExcitation}
  \end{figure}

  In the preceding sections we have focused on the ultralight $c$-axion that can be thought of as the integral of the RR two-form $C_2$ over a sphere between the two throats. Similarly, we can define a $b$-axion by integrating the NS two-form $B_2$ instead, 
  \begin{equation}
   b\equiv \frac{1}{2\pi \ap} \int_{S^2}B_2\, .
  \end{equation}
  By the same arguments as before (see \eqref{eq:LocalFlux}) a non-vanishing field excursion induces a pair of $H_3$ flux/anti-flux on the portions of the $\mathcal{B}$-cycle that reach down into the two throats. Now, in the vacuum the $\mathcal{B}$-cycle is already filled with quantized $H_3$-flux,
  \begin{equation}
   K\equiv K_1+K_2\equiv \frac{1}{(2\pi)^2\ap}\left(\int_{\mathcal{B}_1}H_3 +\int_{\mathcal{B}_2}H_3\right)\, .
  \end{equation}
  Here $\mathcal{B}_1$ and $\mathcal{B}_2$ are the three-chains that reach into the respective throats and are bounded by the sphere between the throats, so that $\mathcal{B}=\mathcal{B}_1+\mathcal{B}_2$. Clearly the continuous field excursion of the $b$-axion does not change the quantized flux integer $K$. However, by Stokes' theorem it \textit{does} change the relative flux distribution,
  \begin{equation}
  K_1\longrightarrow K_1+\frac{b}{2\pi}\, ,\quad K_2\longrightarrow K_2-\frac{b}{2\pi}\, .
  \end{equation}
  By definition, $K_1$ and $K_2$ are the (non-quantized) $H_3$ fluxes that reside in the respective throats. Again, treating the local throat deformation parameters $z_{1,2}$ as independent it is clear that the throats can restore supersymmetry by an appropriate adjustment \cite{Giddings:2001yu}:
  \begin{equation}
  |z_{1,2}|\sim \exp\left(-2\pi \frac{K_{1,2}}{g_sM}\right)\,\,\,\longrightarrow \,\,\,\exp\left(-2\pi \frac{K_{1,2}\pm b/2\pi}{g_sM}\right)\, .
  \label{eq:bbackreaction}
  \end{equation}
  Thus, the discussion of the previous section applies also to the $b$-axion if one replaces the phases of the local deformation parameters by $\log |z_i|$. In other words, while the $c$-axion rotates the throats against each other, the $b$-axion makes one throat longer and the other shorter (see Fig.~\ref{fig:BExcitation}).

  Expanding on the above, we can now comment on the interesting difference between the 5d and 4d perspective on excitations of $\int_{S^2} B_2$ in the local throat. This will in particular facilitate comparison with the related discussion in~\cite{Buratti:2018xjt}.

\begin{figure}[ht]
  	\centering
  	\includegraphics[width=0.5\textwidth,keepaspectratio]{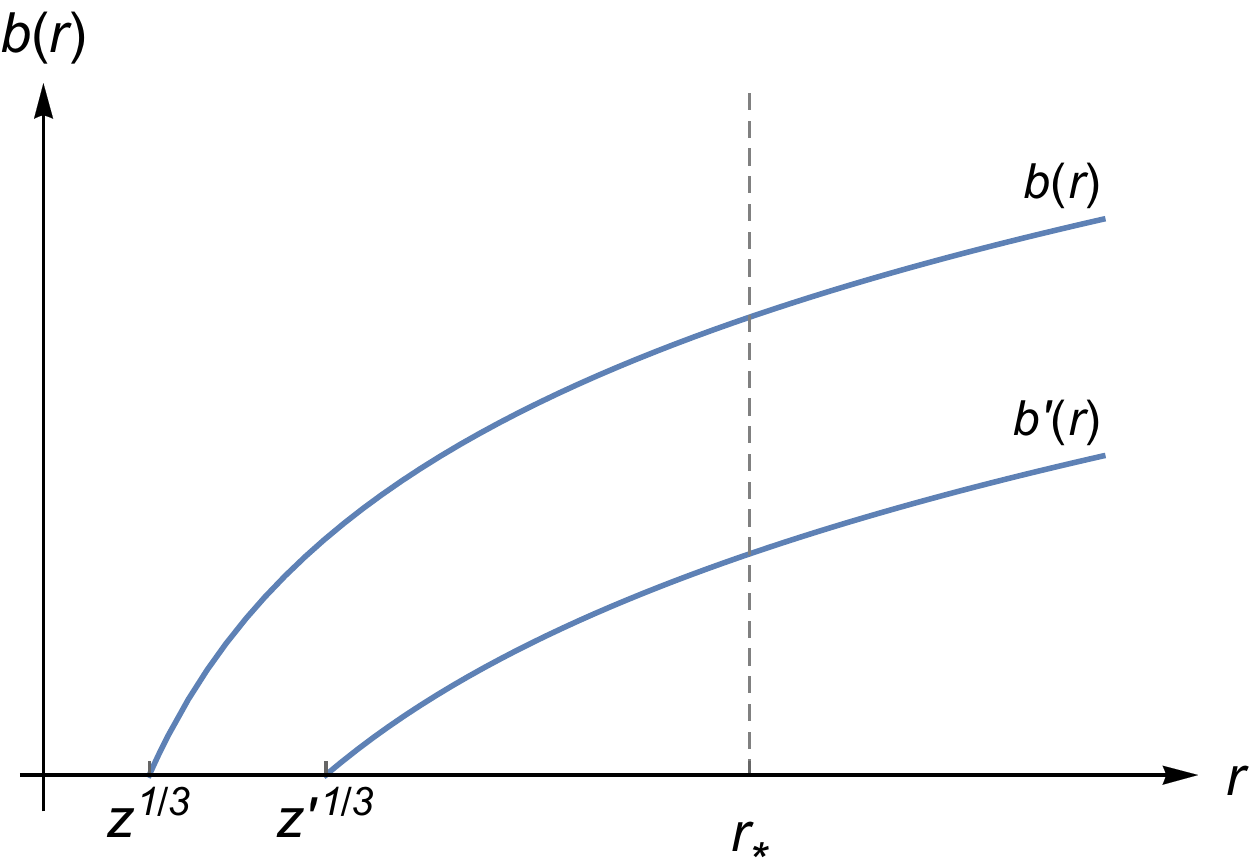}
  	\caption{Two solutions $b(r)\sim\int B_2$ to the supergravity equations of motion in the Klebanov-Tseytlin throat. The 4d flat direction of the potential corresponds to a change of solution.}
  	\label{fig:BProfile}
  \end{figure}

  We start by noting that, instead of the 4d field $b$ defined by an integral inbetween the two throats, one may also consider the 5d field $b(r) \sim \int_{S^2} B_2$ within either of the throats. Here $r$ is the radial location of the relevant $S^2$.\footnote{In order to avoid introducing additional symbols, by a slight abuse of notation, we will denote by $b(r)$ the 5d mode while $b$ without radial argument denotes the 4d mode.} Away from the tip of the throat and from the bulk CY, where the 5d language can be used, one has a continuum of solutions to the 10d supergravity equations~\cite{Klebanov:2000nc}. In particular, there is a continuum of solutions for $b(r)$, parametrized by $z$ via the boundary condition $b\left(r=\abs{z}^{1/3}\right)=0$. Two such solutions are plotted in Fig.~\ref{fig:BProfile}. The relevant 5d equation for a static solution is, symbolically,
  \begin{equation}
   \left[\partial_r^2-f(r)\partial_r-m^2(r)\right]b(r)=0\,,
  \end{equation}
  where $m(r)$ is the 5d mass. One immediately sees that the non-trivial (non-constant) profile of $b(r)$ is enforced by the non-zero $m^2(r)$. This non-zero potential comes from the CS term $\int F_5\wedge B_2\wedge F_3$. One might be concerned that such a potential clashes with the presence of an ultralight (in the present approximation massless) 4d mode of $b$. But such concerns are unfounded: Indeed, the 4d flat direction parameterized by $b$ is also present in the 5d description, in spite of the non-zero 5d potential. It corresponds to a change of the whole $b(r)$-profile within the available continuum of solutions accompanied by a change of boundary condition, e.g. $z\to z'$, cf.~Fig.~\ref{fig:BProfile}. The variable $z$ is, of course, not accessible to a local observer in 5d at some fixed position $r_*$. 
  
  Finally, when gluing together throats to create a double throat one expects a small potential for the $b$-axion due to the misalignment of the \textit{magnitudes} of the two deformation parameters, compare the discussion of Sect.~\ref{sec:CYBreakingPotential} in the $C_2$-case. We will make a quantitative statement in the next section.

 \section{Four-Dimensional SUGRA Completion}
 \label{sec:4dSugra}
 
 So far we have discussed how the $c$-axion backreacts on the phases of the local deformation parameters of the throats. In this section we propose a completion of the model in the language of 4d supergravity. The $C_2$-axion pairs with the analogous $B_2$-axion into a complex field $\mathcal{G}=c-\tau b$.  The $b$-axion backreacts on the \textit{magnitudes} of the deformation parameters in a way that is analogous to the backreaction of the $c$-axion on their phases.
 
 \subsection{Counting Moduli Through the Conifold Transition}
 
 Throughout Sect.~\ref{sec:Upshot} we have focused on the case of two $S^3$-cycles related in homology, i.e.~$[\mathcal{A}_1] = [\mathcal{A}_2]$. In general we denote by $n$ the number of collapsing three-spheres $\mathcal{A}^i$, $i=1,...,n$ and by $m$ the number of homology relations between them $\sum_{i=1}^{n}p_i^I[\mathcal{A}^i]=0$, $I=1,...,m$.
 
 Before fluxes are turned on and orientifold projections are imposed the physical degrees of freedom assemble into $\mathcal{N}=2$ multiplets. The $n-m$ complex structure moduli $z^i$ are the scalar components of $n-m$ vector multiplets. The $z^i$ parametrize the Coulomb branch of the gauge theory. Whenever some of the three-cycles shrink to zero size, charged hypermultiplets (\textit{Strominger black holes}) become massless and have to be `integrated in' \cite{Strominger:1995cz}. These can be thought of as D3-branes that wrap the shrunken cycles. At the origin of the Coulomb branch there are hence $n$ massless charged hypermultiplets and $n$ singular nodes have developed in the CY threefold. There exists an $m$-dimensional Higgs-branch where the singular nodes are resolved into $m$ (homologically independent) $\mathbb{P}^1$'s \cite{Greene:1995hu}. On this branch, the $n-m$ vector multiplets eat $n-m$ black brane hypermultiplets and become massive. Geometrically speaking this is the resolution of the conifold \cite{Candelas:1989js,Candelas:1988di}.
 
 In the $\mathcal{N}=1$ flux compactification that we are considering the tips of the conifolds become strongly red-shifted. Moreover backreaction of fluxes ensures that even at tiny complex structure the $S^3$'s stay at finite size so that the Strominger black holes play no role. However, since the deformed and the resolved conifold differ only by their strongly red-shifted tip geometries we expect to recover some remnant of the resolved phase of the conifold theory in the light spectrum. As outlined in Sect.~\ref{sec:Upshot} we expect the `local complex structures' to decouple from one another so that all the $n$ local deformation parameters $z_1,...,z_n$ become equally light. In other words there are $m$ additional light geometric modes. Moreover, on the resolved side of the transition there would be $m$ massless axion modes. Since the obstruction for them to be massless is also localized at the tips of the conifolds where the would-be two-cycles collapse we also expect $m$ complex light axionic modes $\mathcal{G}^I$. As we will argue in the next section, these modes indeed appear quite naturally in the discussion of the flux superpotential.
 
 \subsection{The Thraxion Superpotential}
 \label{sec:Sugra}
 
 In this section we make a proposal for the 4d supergravity completion of the Lagrangians \eqref{eq:eff.Lagr1} and \eqref{eq:eff.Lagr2} for a general number of throats $n$ with $m$ homology relations among the shrinking cycles. Throughout this section we work in units $M_\Pl=1$.
 
 \subsubsection{The GVW Superpotential of a Multi Conifold System}
 As a starting point we consider the GVW flux superpotential for a multi conifold system. All the necessary ingredients are derived in ref. \cite{Greene:1996cy} and summarized in Appendix \ref{Appendix:BackgroundOnConifolds}. We choose to treat the redundant set of the $n$ complex structure parameters $z_i$ associated with the $n$ vanishing cycles $\mathcal{A}^i$ democratically, and impose the $m$ \textit{CY conditions} via a set of Lagrange multipliers $\lambda_I$, $I=1,..,m$. The superpotential reads
 \begin{equation}\label{eq:n-conifold-superpotential1}
 W(z) =\sum_{i=1}^{n}\left(M_i\frac{z_i}{2\pi i} \log(z_i)+M_ig^i(z)-\tau K^i z_i\right)+\sum_{I=1}^{m}\lambda_I P^I+\hat{\hat{W}}_0(z) \, .   
 \end{equation}
 The $m$ homology relations among the vanishing cycles $\sum_{i=1}^{n}p_i^I\mathcal{A}^i=\del \mathcal{C}^I$, $I=1,...,m$ lead to the following $m$ CY conditions for the $z_i\equiv \int_{\mathcal{A}^i}\Omega$,
 \begin{equation}
 0\overset{\diff\Omega=0}{=}\int_{\mathcal{C}^I}\diff\Omega =\int_{\del\mathcal{C}^I}\Omega = \sum_{i=1}^{n}p^I_i \int_{\mathcal{A}^i}\Omega=    \sum_{i=1}^{n}p^I_i z_i \equiv  P^I\, ,\quad I=1,...,m\, .
 \end{equation} 
 In this language, the $m$ CY conditions $P^I=0$ are equivalent to the F-term equations of the Lagrange multipliers $\lambda_I$, $\del_{\lambda_I}W\overset{!}{=}~0$. For details we refer the reader to App.~\ref{Appendix:BackgroundOnConifolds}\footnote{Note that we restrict ourselves to regions in complex structure moduli space close to the conifold transition point, where all throats degenerate simultaneously, compare the discussion in Sect.~\ref{sec:GeometricBackground}. This might be more restrictive than is needed for our analysis: If the matrix $p_i^I$ is block-diagonal, we can separate the multi throat system into smaller multi throats whose deformations are independent of one another. In this case we can go through a conifold transition by local degeneration of the throats of a smaller system. Even away from the trivial case of multi throats factorizing, one might be able to achieve small thraxion masses by `freezing' individual throats with larger deformation $z$. Given a multi throat with some large $z$'s one has to check the thraxion potential as proposed in this section for flat directions. We leave a more thorough analysis of this possibility for future work.}.
 
 Here, the $M_i$ and $K^i$ are the flux numbers associated to the $\mathcal{A}$- and $\mathcal{B}$-cycle of the $i$-th throat, and the holomorphic function $\hat{\hat{W}}_0(z)$ denotes contributions to the flux superpotential from other cycles. The $M_i\in \Z$ cannot all be chosen independently but must comply with the $m$ homology conditions 
 \begin{equation}\label{eq:rank-condition}
 \sum_{i=1}^{n}p^I_i M_i=0\, ,\quad I=1,...,m\, .
 \end{equation}
 The $K^i$ can be chosen independently but there is an $m$-fold redundancy in their definition because we may transform $K^i\longrightarrow K^i+\sum_I \alpha_I p^I_i $ for any $\alpha\in\C^m$ leaving the superpotential invariant upon imposing the constraint equations\footnote{The $n-m$ physical $H_3$ flux quantization conditions can be stated as $K^a - \sum_{I=1}^m p^{I}_a K^{n-m+I} \in \Z$, $a=1,\ldots,n-m$. This is because we can always choose the first $n-m$ of the shrinking cycles to correspond to integral basis elements $[\mathcal{A}^1],...,[\mathcal{A}^{n-m}]$ in homology. The Lagrange constraints can be stated as $0=P^I=\sum_{a=1}^{n-m}p^I_a z_a+z_{n-m+I}$, i.e.~$z_{n-m+I}=-\sum_{a=1}^{n-m}p^I_a z_a$. In the superpotential the terms that multiply $z_1,...,z_{n-m}$ are given by the above combination of $K^i$ and correspond to the integer flux numbers on the cycles $\mathcal{B}_1,...,\mathcal{B}_{n-m}$. Alternatively, one may demand the sufficient but not necessary conditions that $K^i \in \Z$ for $i=1,\ldots,n$. In this more restrictive but democratic formulation the $i$-th throat carries $K^i$ units of flux. We can still reach all possible integer values for flux numbers on the cycles $\mathcal{B}_a$.}. Furthermore, there are $n$ unknown functions $g^i(z)$ defined on complex structure moduli space that are holomorphic near the origin.
 
 The K\"ahler potential is given by
 \begin{equation}\begin{split}\label{eq:Kahlerpot_cs1}
 K_\text{cs}(z_i,\bar{z}_i)=&-\log\left(-i \int \Omega \wedge \bar{\Omega} \right)=-\log\left(ig_K(z)-i\overline{g_K(z)} + \sum_{a=1}^{n-m}i\bar{z_a}G^a+c.c.\right)\\
 =&-\log\left(ig_K(z)-i\overline{g_K(z)}+\sum_{i=1}^{n}\left[\frac{|z_i|^2}{2\pi}\log(|z_i|^2)+i\bar{z}_ig^i(z)-iz_i\overline{g^i(z)}\right]\right)\, ,
 \end{split}\end{equation}
 where the holomorphic function $g_K(z)$ encodes contributions from other cycles. We would like to stress that despite the fact that we have written the unknown functions $g^i,g_K$ and $\hat{\hat{W}}_0$ as functions of all the $z_i$, $i=1,...,n$, knowledge of the periods of the various cycles (and the flux quanta) only determines their behavior \textit{along} complex structure moduli space and not beyond.
 
 \subsubsection{The Thraxion as a Stabilizer Field}
 
 We are now ready to formulate a proposal for the thraxion superpotential. First, we note the following. By expanding the Lagrange multiplier terms, one may rewrite the superpotential \eqref{eq:n-conifold-superpotential1} as
  \begin{equation}
  W(z) =\sum_{i=1}^{n}\left(M_i\frac{z_i}{2\pi i} \log(z_i)+M_ig^i(z)+\left[-\tau K^i+\sum_{I=1}^{m}\lambda_I p_i^I\right] z_i\right)+\hat{\hat{W}}_0(z) \, .   
  \end{equation}
 One observes immediately that the combinations $\sum_{I=1}^{m}\lambda_Ip^I_i$ can be interpreted as an \textit{additional}, unquantized contribution to the complex three-form flux $G_3=F_3-\tau H_3$ on the (local portion $\tilde{\mathcal{B}}^i$ of the) $\mathcal{B}$-cycle of the $i$-th conifold. But we know that such a flux is detected by a boundary integral
 \begin{equation}
 \hat{\mathcal{G}}_i \equiv c_i - \tau b_i\equiv \frac{1}{2 \pi \ap}\int_{S^2|_{\text{$i$-th throat}}}(C_2-\tau B_2) = \frac{1}{2 \pi \ap}\int_{\tilde{\mathcal{B}}^i}(F_3-\tau H_3)
 \end{equation} 
 over the $S^2$ at the top of the $i$-th throat. Crucially, the variables $\hat{\mathcal{G}}_i$ define axionic field excursions as measured near the entrance of the $i$-th throat.
 
\begin{figure}[t]
 	\centering
 	\def\svgwidth{0.5\textwidth}
 	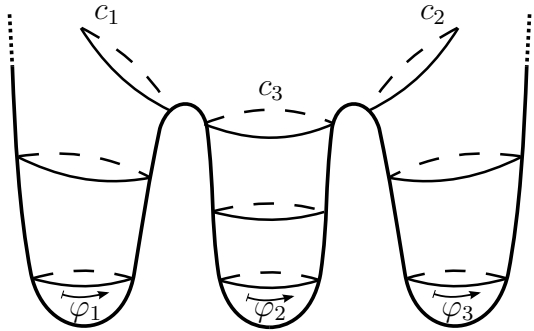
 	\caption{A cartoon of the $\mathcal{B}$-cycle of the triple throat, $n=3$ and $m=2$. The local $c$-axion excursions $c_1,c_2$ and $c_3$, $c_i = \text{Re}\,{\hat{\mathcal{G}}_i}$, must be chosen so that no overall flux is generated on $\mathcal{B}$, i.e.~$0\overset{!}{=}c_1+c_2+c_3$\,\protect\footnotemark~or rather $\sum_i \hat{\mathcal{G}}^i = 0$.}
 	\label{fig:MultiThroat_TripleThroat}
 \end{figure}

 We would like to interpret (a subset of) these as light physical degrees of freedom. This is motivated by the fact that there are $m$ light axions on the other side of the conifold transition that correspond to the integrals of $C_2-\tau B_2$ over the independent resolution $2$-cycles. Indeed, the counting is correct. A consistent axionic field excursion must not induce any overall flux on \textit{any} of the global $\mathcal{B}$-cycles (see Fig.~\ref{fig:MultiThroat_TripleThroat}). There are hence $n-m$ \textit{no-flux} conditions, one for each linearly independent $\mathcal{B}$-cycle, leaving only $m$ physical axions. These can be parametrized as $\hat{\mathcal{G}}^i=\sum_{I=1}^{m}p_i^I\mathcal{G}_I$ and we are led to the following conjecture: 
 \begin{center}
 	\begin{tabular}{ m{9cm} }
 		\textit{The Lagrange multipliers $\lambda_I$ must be promoted}
 		\textit{to $m$ light axionic degrees of freedom, $\lambda_{I}\longrightarrow \frac{\mathcal{G}_I}{2\pi}$. Moreover, the $z^i$ are promoted to $n$ physically independent degrees of freedom.} 
 	\end{tabular}
 \end{center}
 The normalization factor $2\pi$ is chosen such that locally in the $i$-th throat a shift of axionic field excursion $\mathcal{G}_I$ by $2 \pi$ (or $2\pi \tau$) for some $I$ is indistinguishable from an increase of the $F_3$-flux (respectively $H_3$-flux) on the $\mathcal{B}$-cycle of the throat by an integer amount $p^i_I$.
 
 \footnotetext{Compare this to Fig.~\ref{fig:Upshot_Fluxes}: The `no-flux' condition in the double throat setup amounts to $c_1 = -c_2$. The two axions $c_1$ and $c_2$ are actually identified, up to a sign due to different orientation of the two-sphere in the definition. This is why we only had one axion $c$ to begin with in the 10d analysis of Sect.~\ref{sec:Upshot}.}
 
 Thus, our propoal for the superpotential is
 \begin{equation}\label{eq:nm-throat-superpotential}
 W=\sum_{i=1}^{n}\left(M_i\frac{z_i}{2\pi i} \log(z_i)+M_ig^i(z)-\tau K^i z_i\right)-\sum_{I=1}^{m}\frac{\mathcal{G}_I}{2\pi} P^I+\hat{\hat{W}}_0(z)\, .
 \end{equation}
 We find it interesting to note that the axions $\mathcal{G}^I$ now serve as the \textit{stabilizer fields} for the combinations of the local deformation parameters that break the $m$ CY conditions $P^I=0$, $I=1,...,m$. This form of the dynamical thraxion superpotential is fairly unique in that it preserves the set of \textit{discrete} shift-symmetries
 \begin{equation}
 z_i\longrightarrow z_i e^{\frac{2\pi i}{M_i} \sum_I\eta_I p_i^I}\, ,\quad \mathcal{G}_I\longrightarrow \mathcal{G}_I+2\pi \eta_I\, , \quad \forall \eta\in\C^m \, : \sum_I \eta_I p^I_i\in M_i\Z \quad  \forall i\, .
 \end{equation}
 Our proposal for the K\"ahler potential is 
 \begin{equation}
 K(\mathcal{G}_I,\bar{\mathcal{G}}_{\bar{I}},T,\bar{T},z,\bar{z})=K_1(\mathcal{G}_I-\bar{\mathcal{G}}_{\bar{I}},T+\bar{T})+K_\text{cs}(z,\bar{z})\, ,
 \end{equation}
 where $K_\text{cs}$ is the K\"ahler potential \eqref{eq:Kahlerpot_cs1} and $K_1$ is the K\"ahler potential of the $m$ axions (and K\"ahler moduli $T$) on the other side of the conifold transition as derived in \cite{Grimm:2004uq}
 \begin{equation}
 K_1 = K_0 -3 \log\left( T+ \bar{T} - \frac{3 i}{4(\tau-\bar{\tau})} \kappa_{1IJ}(\mathcal{G}-\bar{\mathcal{G}})^I(\mathcal{G}-\bar{\mathcal{G}})^J \right)\,,
 \label{eq:Kahlerpot}
 \end{equation}
 where $K_0$ contains a constant part and the K\"ahler potential of the axio-dilaton and $\kappa_{1IJ}$ are triple intersection numbers.\footnote{\label{footnote:no-scale}For ease of exposition we have stated the K\"ahler potential for the case $h_{1,1}^+=1$. For $\tau = \text{const.}$ the K\"ahler potential fulfills the no-scale relation $(\partial_{T} K_1) (K_1^{-1})^{T\bar{T}} (\partial_{\bar{T}} K_1) = 3$, and $(K_1^{-1})^{\mathcal{G}^I\bar{X}^{\bar{j}}}\del_{\bar{X}^{\bar{j}}}K_1=0$ where $X^i = (T,\mathcal{G}^I)$.}
 
 We expect \eqref{eq:Kahlerpot_cs1} and \eqref{eq:nm-throat-superpotential} to hold even when we break the CY condition $P^I\neq 0$ with the important subtlety that the domain of the holomorphic functions $g^i,g_K$ and $\hat{\hat{W}}_0$ must be extended \textit{beyond} complex structure moduli space. We find it reasonable to expect that such an extension \textit{exists} although even full knowledge of the CY periods would not determine their behavior away from the moduli space. The detailed form of these functions will be of no importance in what follows. Moreover, we expect that using the potential $K_1$ gives an excellent approximation because the kinetic term of the axions is dominated by contributions from the UV where the deformation or resolution of the conifold plays but a tiny role. Note, that the behavior of the kinetic terms of the complex structure moduli is dominated by the logarithmic terms in \eqref{eq:Kahlerpot_cs1}. In particular, the functions $g^i(z)$ and $g_K(z)$ contribute to kinetic terms only at sub-leading order.
 
 Since we are interested in small $z_i$ we Taylor-expand
 \begin{align}\label{eq:thresholdcoeff}
 g^i(z)=&g_0^i+\sum_{j=1}^{n}g^{ij}_{1} z_j+\mathcal{O}(z^2)\, ,\quad \hat{\hat{W}}(z)=g_{W,0}+\sum_{i=1}^{n}g_{W,1}^iz_i+\mathcal{O}(z^2)\, ,\nonumber\\
 g_{K}(z)=&g_{K,0}+\sum_{i=1}^{n}g_{K,1}^i z_i+\mathcal{O}(z^2)\, .
 \end{align}
 This should really be understood as a Taylor expansion in $n$ \textit{independent} variables $z_i$ and makes our conjectured extension of the domain of these functions beyond the complex structure moduli space manifest.
 
 We absorb all $\mathcal{O}(z^0)$ terms in the superpotential in the definition $\hat{W}_0\equiv g_{W,0}+\sum_{i=1}^{n}M_i g^i_0$. The coefficients in \eqref{eq:thresholdcoeff} should all be viewed as independent of the flux quanta that thread the cycles of the multi throat system, and only $(g_{W,0},g_{W,1}^i)$ depend on fluxes on other cycles.
 
 \subsubsection{The Double Throat: \texorpdfstring{$n=2$}{n=2}, \texorpdfstring{$m=1$}{m=1}}
 
 It is clear that to obtain the effective superpotential for the $\mathcal{G}$-fields we should integrate out the local deformation parameters. Before we discuss this in full generality it is instructive to first consider the simplest case of the double throat, i.e.~$n=2$ and $m=1$. There are two $\mathcal{A}$-cycles $\mathcal{A}^1$ and $\mathcal{A}^2$ and we choose the homology relation to be $\mathcal{A}^1\sim \mathcal{A}^2$. Hence, there are two deformation parameters $z_1$ and $z_2$ and one axion $\mathcal{G}$. For ease of exposition we assume that of all the coefficients defined in \eqref{eq:thresholdcoeff}, only $g_{W,0}$ and $g_{K,0}$ are non-vanishing, in other words, we choose $\hat{W_0}$ as well as all non-logarithmic terms in the K\"ahler potential to be constant. In doing so we accept an $\calO(1)$ error in all expressions, in particular in the resulting superpotential $W_\text{eff}$ for $\mathcal{G}$. This simplifying assumption will be dropped when we generalize the discussion to the multi throat case in Sect.~\ref{sec:Multithroat}.
 
 We must set $M_1=M_2\equiv M$ due to the homology relation between the shrinking cycles, and we choose $K^1\equiv K/2  \equiv K^2$ which results in flux $K^1 + K^2 = K$ on the $\mathcal{B}$-cycle. All choices of the pair $(K^1,K^2)$ that satisfy $K^1+K^2=K$ are physically equivalent to this choice and can be brought back to the symmetric choice via a linear redefinition of $\mathcal{G}$ (compare the discussion below \eqref{eq:rank-condition}). The superpotential takes the form
 \begin{equation}
 \begin{split}
 \label{superpotential}
 W(z_1,z_2,\mathcal{G})=&\sum_{i=1}^{2}\left(\frac{z_i}{2\pi i}\log (z_i)M  -\frac{1}{2}K \tau z_i\right)-\frac{\mathcal{G}}{2\pi} (z_1-z_2)
 +\hat{W}_0+\calO(z_i^2)\, .
 \end{split}
 \end{equation}
 First, the F-terms $F_{z_i}$ are given by
 \begin{equation}
 \begin{split}
 D_{z_i}W&=\del_{z_i}W+(\del_{z_i}K)W=\frac{\log (z_i)+1}{2\pi i}M-\frac{K}{2}\tau \mp \frac{\mathcal{G}}{2\pi}+\calO(z_i)\\
 &= \frac{M}{2\pi i}\left(\log(z_i/z_0)\mp i\mathcal{G}/M\right)+\calO(z_0)\, ,
 \end{split}
 \end{equation}
 where
 \begin{equation}
 z_0= e^{-1}\exp\left(2\pi i\tau\frac{K/2}{M}\right)+\mathcal{O}\left(e^{-4\pi \frac{K/2}{g_sM}}\right)= \mathcal{O}\left(e^{-2\pi \frac{K/2}{g_sM}}\right)\, .
 \end{equation}
 As usual, with $K/2> g_sM$ one obtains $|z_0|\ll 1$ with universal dependence on the flux numbers. Following Sect.~\ref{sec:Upshot} we may integrate out the local deformation parameters, which yields 
 \begin{equation}
 z_1=z_0e^{i\mathcal{G}/M}\, ,\quad z_2=z_0e^{-i\mathcal{G}/M}\, .
 \end{equation}
 The effective superpotential for the axion $\mathcal{G}$ reads
 \begin{equation}
 \begin{split}
 W_\text{eff}&(\mathcal{G})=2\epsilon \left(1-\cos(\mathcal{G}/M)\right)+W_0+\calO(z_0^2)\, ,\\
 \text{with}\quad \epsilon&\equiv M\frac{z_0}{2\pi i}\, ,\quad \text{and} \quad W_0\equiv \hat{W}_0-\frac{z_0}{\pi i}M\, .
 \label{eq:EffSuperPot}
 \end{split}
 \end{equation}
 This is the expression we were after. Crucially, it is consistent with the results of Sect.~\ref{sec:Upshot}: Using the K\"ahler potential for the K\"ahler moduli and $\mathcal{G}$-axions \eqref{eq:Kahlerpot}, one may show that $V(\mathcal{G},\bar{\mathcal{G}})\propto |\partial_\mathcal{G} W(\mathcal{G})|^2$ \footnote{Where we use that $K_1$ is of no-scale type.}. If we restrict to $\mathcal{G}=c\in \R$, we reproduce the periodic potential \eqref{eq:vc} with the correct scaling $\abs{\epsilon}^2 \sim \abs{z_0}^2 \sim  w_\IR^6$.
 
 Note that because we have made use of the \textit{unwarped} K\"ahler potential we do not reproduce the correct mass-scale of the local deformation parameters $z_i$. Here, this is of no importance because all degrees of freedom that are related to strongly warped regions are integrated out supersymmetrically. In particular, the potential energy induced by a non-vanishing field excursion of the field $\mathcal{G}$ receives its dominant contributions from the bulk CY where warping plays no role. Because in going from weak to strong warping, the solutions of the complex structure F-terms are left invariant \cite{Gukov:1999ya}, and because the $z_i$ are parametrically heavier than $\mathcal{G}$ even when the appropriate red-shift factors are introduced in the scalar potential, this procedure is justified.
 
 We are now ready to expand on the conclusions we have drawn in Sect.~\ref{sec:Upshot}. First of all,
 the kinetic term of the full complex field $\mathcal{G}$ lives in the bulk. This implies that the mass$^2$ of $\mathcal{G}$ is of order $|z_0|^2\ll 1$. Since the K\"ahler potential is independent of $\text{Re}(\mathcal{G})$ a discrete shift-symmetry $\mathcal{G}\longrightarrow \mathcal{G}+ 2\pi M$ is manifest\footnote{In an exact no-scale background the scalar potential even has periodicity $\pi M$. However, any no-scale breaking effects will break it to the periodicity of the superpotential.}, while the IR superpotential breaks the shift-symmetry corresponding to $\text{Im}(\mathcal{G})$ completely.
 
 While in principle the target space distance traversed by $\text{Im}(\mathcal{G})$ can be made large, the scalar potential grows exponentially as a function thereof as is common for saxionic directions in field space. In particular, this direction in field space is of little use for (slow-roll) inflation. There is a \textit{critical} field excursion $|\text{Im}(\mathcal{G})_\text{crit}|\lesssim 3 M \log (w_{\IR}^{-1})$ beyond which one side of the double throat is entirely pulled up into the bulk CY, $z_1 \sim 1$ or $z_2 \sim 1$. Near this field excursion we no longer know the form of the potential because we work to lowest order in $|z_1|,|z_2|$. Moreover, there is a tower of warped KK-modes with masses that scale as 
 \begin{equation}
 m_n^2\sim n^2 w_{\IR}^2\exp\left(- 2|\text{Im}(\mathcal{G})|/3 M\right)\, ,
 \end{equation}
 where the warp factor $w^2 \sim \abs{z_i}^{2/3}$ now depends on $\text{Im}(\mathcal{G})$. Since these modes have been integrated out, the ratio $\Lambda/m_{\mathcal{G}}$ of the cutoff of the $\mathcal{G}$-EFT (i.e.~the smallest KK-mass) over the mass-scale of $\mathcal{G}$, comes down as 
 \begin{equation}
 \Lambda/m_\mathcal{G} \sim w_{\IR}^{-2}e^{-\frac{4}{3M}|\text{Im}\,\mathcal{G}|}\leq w_{\IR}^{-2}\exp\left(-\frac{\phi_b}{M_\Pl}\right)\, ,
 \end{equation}
 at large field excursion, consistent with a \textit{distance conjecture} \cite{Ooguri:2006in,Klaewer:2016kiy,Blumenhagen:2017cxt,Grimm:2018ohb,Heidenreich:2018kpg,Lee:2018urn,Lee:2018spm,Grimm:2018cpv}. Here $\phi_b$ measures the canonical field distance from the origin along the imaginary $\mathcal{G}$-axis\footnote{See App.~\ref{Appendix:AxionDecayConstant} for the conversion rule between $\mathcal{G}$ and the canonical distance in field space $\phi_b$. At any point in field space, $g_{\mathcal{G}\overline{\mathcal{G}}}<M_\Pl^2/M^2$. Hence, $\phi_b=\int_{0}^{\text{Im}\,\mathcal{G}}\diff \, \text{Im}\,\mathcal{G}' \sqrt{g_{\mathcal{G}\overline{\mathcal{G}}}}<\text{Im}\,\mathcal{G}M_\Pl/M$.}.
 
 At strong warping the maximal allowed field excursion is $|\text{Im}(\mathcal{G})_\text{max}|\sim \frac{3}{2} M \log(w_{\IR}^{-1})$ before the 4d EFT description breaks down. Near this field excursion, a large fraction of the reservoir of fluxes of one of the throats has been transferred to the other one and the mass scale of the $\mathcal{G}$-field is of the same order as the warped KK-scale of the longer throat. At this point, contributions to the scalar potential from non-vanishing $F$-terms $D_{z_i}W$ start to play a significant role, or from a 10d point of view, the potential energy sinks down into the longer throat.
 
 \subsubsection{The General Multi Throat}
 \label{sec:Multithroat}
 
 For general $m$ and $n$ that satisfy $n-m>0$ homology relations, there are $n$ local deformation parameters $z_1,...,z_n$ that have to be integrated out. We are left with an effective supergravity theory of $m$ axions $\mathcal{G}^I$. The computational steps are analogous to the double throat case that was laid out in detail. Hence, we only state the effective axion superpotential
 \begin{equation}
 W_\text{eff}(\mathcal{G}^I)=-\sum_{i=1}^{n}\epsilon_i e^{i\sum_{I=1}^{m}p^i_I \mathcal{G}^I/M_i}+\hat{W}_0\, ,
 \label{eq:EffSuperPotMulti}
 \end{equation}
 and we have defined
 \begin{align}
 \epsilon_i\equiv \frac{M_i}{2\pi i}z_{0,i}(1-2\pi \bar{\tilde{g_0}}^i\hat{W_0}/(a M_i))\, ,\quad \tilde{g}_0^i\equiv g_0^i-\overline{g_{K,1}^i}\, , \quad a\equiv -2 \text{Im}(g_{K,0})\, ,
 \end{align}
 and
 \begin{equation}
 z_{0,i}= e^{-1}e^{-\frac{2\pi i}{M_i} \left(\sum_j M_j g_{1}^{ji}+g_{W,1}^i+i\bar{\tilde{g_0}}^i\hat{W_0}/a\right)}\exp\left(2\pi i\frac{K^i\tau}{M_i}\right)+\mathcal{O}\left(e^{-4\pi \frac{K^i}{g_sM_i}}\right)\, .
 \end{equation}
 It is important to note that the $z_{0,i}$ as defined above can in general \textit{not} be interpreted as the values of the local deformation in the vacuum. The physical local deformation parameters are given by
 \begin{equation}
 z_{\text{ph.},i}\equiv z_{0,i}\exp\left(i\sum_I p^i_I \mathcal{G}^I/M_i\right)\, ,
 \end{equation}
 where in the vacuum the $\mathcal{G}^I$ need not vanish in general.

 \subsection{Comments on the \texorpdfstring{$b$}{b}-Axion}\label{sec:b-axion}
 
 In the above supergravity completion we have `complexified' the $c$-axion by pairing it with the analogous $b$-axion. We have outlined the 10d backreaction of the $b$-axion already in Sect.~\ref{sec:b-axion0}. Now that we have addressed the scalar potential of the $b$-axion quantitatively, in this section we would like to comment on a potential worry and how to resolve it: We recall that the effect of a non-vanishing field excursion of the $b$-axion is the creation of a pair of fluxes of the NS field strength $H_3$. Since both throats are filled up with $H_3$-fluxes already in the vacuum one should think about this process more properly as a transfer of $H_3$-flux from one throat to the other. Since the magnitudes of the local deformation parameters (and the associated hierarchy) are set by the ratio of local $H_3$-flux (on the $\mathcal{B}$-cycle) and $F_3$-flux (on the $\mathcal{A}$-cycle) it is clear that these will backreact when the $H_3$-fluxes are redistributed, see \eqref{eq:bbackreaction}.
 
 However, it is also clear that when the local $H_3$-fluxes are changed, the circumference of the throat at the UV end is affected strongly. This is because it is set by the total D3-charge that is stored in the throat which is itself proportional to the amount of $H_3$-flux \cite{Klebanov:2000hb}, compare Fig.~\ref{fig:BExcitation}. Hence, naively one might worry that such considerable change at the UV ends of the throats could lead to a large potential energy. One may convince one-self that this is not the case as follows. Starting from the supersymmetric situation we can redistribute a small amount of fluxes from one throat to the other, so that throat $A$ has $\delta$ units of $H$-flux more than throat $B$. We can now proceed to convert the extra fluxes into a number of D3-branes by going through the Kachru-Pearson-Verlinde (KPV) transition \cite{Kachru:2002gs}. From the UV perspective this process is only detected by a change in the throat complex structure which is a tiny perturbation far from the tip of the throat. Now we are back to an even flux distribution with a number of mobile D3-branes. These can be moved out of the throat at no cost in energy so the situation with the mobile branes should be a vacuum again. In other words, the redistribution of fluxes creates an energy density that is \textit{only} due to the misalignment of local deformation parameters and the change of size of the throats at their UV ends does not generate an extra contribution to the potential. We reiterate that the situation is analogous to the backreaction of the $c$-axion with the phases of the local deformation parameters replaced by the logarithms of their magnitude.
 
 Finally, note that in the K\"ahler potential \eqref{eq:Kahlerpot} the $b$-axion appears explicitly, while the approximate $c$-axion shift symmetry is manifest. One might suspect that the small scale of the $b$-axion is therefore accidental due to our use of tree level supergravity. This conclusion would be incorrect: The target space manifold with K\"ahler metric derived from \eqref{eq:Kahlerpot} is shift symmetric also in the $b$-direction \cite{Antoniadis:2015egy,Corvilain:2016kwe}. In general we expect both shift symmetries to be preserved to all orders in the perturbative expansion with explicit breaking only due to the superpotential \eqref{eq:EffSuperPotMulti}. When moreover \textit{non-perturbative} K\"ahler moduli dependent terms are generated in the superpotential such as the ones considered in \cite{Kachru:2003aw}, we expect the $b$-axion mass to be lifted to the scale of K\"ahler moduli stabilization while the $c$-axion can remain parametrically lighter.

   \section{The Axion Potential and the Gauge/Gravity Correspondence}\label{sec:gaugegravity}
   
   We have derived the axion potential via a classical computation within 10d SUGRA, and proposed a 4d SUGRA description that matches it. Since the local throats are believed to have a dual description in terms of KS gauge theories \cite{Klebanov:2000hb}, it is useful to give an alternative derivation of our results on the gauge theory side of the correspondence. The KS gauge theory is a $SU(N+M)\times SU(N)$ gauge theory with (classical) global symmetry group $SU(2)\times SU(2)\times U(1)_R$. It contains matter in bi-fundamental representations $(\square,\overline{\square})$ and $(\overline{\square},\square)$ of the gauge group that transform as doublets under the first respectively second global $SU(2)$ factor. The holomorphic gauge couplings of the two gauge theory factors $\tau_{YM}$ and $\tilde{\tau}_{YM}$ have been argued to be set by \cite{Klebanov:1998hh,Herzog:2002ih}
   \begin{equation}
   \tau_\text{YM}+\tilde{\tau}_\text{YM}=\tau\, ,\quad \tau_\text{YM}-\tilde{\tau}_\text{YM}=-\tau+\mathcal{G}/\pi\mod 2(m-n\tau)\, ,
   \end{equation}
   with $(m,n) \in \Z^2$. The radial running of the $\mathcal{G}$-field together with $\tau=\text{const.}$ matches the RG-running of the gauge theory coupling constants. Throughout this section $\mathcal{G}$ takes values in its suitable fundamental domain.
   
   As the KS gauge theory flows to the infrared, it undergoes repeated steps of Seiberg dualities that reduce the ranks of the gauge groups according to
   \begin{align}
   SU(N_0+M)\times SU(N_0)&\longrightarrow  SU(N_1+M)\times SU(N_1)\nonumber\\
   &\longrightarrow \cdots \longrightarrow  SU(N_k+M)\times SU(N_k)\, ,
   \end{align} 
   with $N_k\equiv N-kM$, $k\in \N$. If we start with $N=KM$, after $K$ steps in the duality cascade the gauge group is $SU(M)$. Since, roughly speaking, it corresponds to the first gauge group factor in $SU(M)\equiv SU(N_K+M)\times SU(N_K)$, its holomorphic scale is given by $\Lambda^{3M}=\mu_{\IR}^{3M}\exp{(2\pi i \tau_\text{YM}(\mu_{\IR}))}=\mu_{\IR}^{3M} \exp(i\mathcal{G})$, where $\mu_{\IR}$ is the infrared scale of the throat and where we make use of the KS dictionary $\tau_\text{YM}\simeq \mathcal{G}/2\pi$. Gaugino condensation leads to an effective ADS-superpotential \cite{Veneziano:1982ah,Affleck:1983mk,Klebanov:2000hb}
   \begin{equation}
   W_\text{eff}(\mathcal{G})=M \Lambda^3\sim M\mu_{\IR}^3\exp(2\pi i \tau_\text{YM}/M)\sim M\mu_{\UV}^3\exp\left(\frac{2\pi i}{M}\left(\tau K +\frac{\mathcal{G}}{2\pi}\right)\right),
   \end{equation}
   where we have used that the IR-scale is related to the UV-scale by $\mu_{\IR}^3=\mu_{\UV}^3\exp(-2\pi \frac{K}{g_sM})$.
   
   The superpotential that we have proposed on the gravity side of the correspondence \eqref{eq:EffSuperPotMulti} indeed takes this form,
   \begin{equation}
   W \propto \sum_{i=1}^{n}M_iA_i \exp\left(\frac{2\pi i}{M_i} \left(\tau K^i+\frac{\hat{\mathcal{G}}^i}{2\pi}\right)\right)+\hat{W}_0\, ,
   \end{equation}
   with $\hat{\mathcal{G}}^i=\sum_{I=1}^{m}p^i_I \mathcal{G}^I$, and\footnote{Note that in \eqref{eq:EffSuperPotMulti} we have set $M_\Pl=1$. Therefore we identify $\mu_\UV\sim M_{\Pl}$.}
   \begin{equation}\label{eq:thresholdcoeff2}
   A_i\equiv \exp\left(-\frac{2\pi i}{M_i}\left(\sum_{j=1}^{n}M_j g^{ji}_1+g_{W,1}^i+i\bar{\tilde{g}}_0^i \hat{W}_0/a\right)+\log\left(1-\frac{2\pi \bar{\tilde{g}}_0^i}{aM_i}\hat{W}_0\right)\right)\, .
   \end{equation} 
   From the gauge theory perspective we should interpret the appearance of the constants $g_1^{ji}, g_{W,1}^i$, $\tilde{g}_0^i$ and $\hat{W}_0$ as a parameterization of threshold corrections near the UV cutoff\footnote{Of course these are in general functions of all other complex structure moduli that do not control the infrared regions of the throats and are frozen at a high scale.}. Indeed, as they are taken to zero the $A_i$ become unity.
   
   It is now obvious that the $M$-fold extension of the periodicity of the $c$-axion is related to gaugino condensation in the KS gauge theory\footnote{Related observations were made in the non-compact flux-less multi-node setting of \cite{Heckman:2007wk,Heckman:2007ub}.} \cite{Klebanov:2000nc,Klebanov:2000hb,Cachazo:2001jy,Herzog:2002ih}: As usual there is a $U(1)_R$ symmetry that is broken to $\Z_{2M}$ by gauge theory instantons. Gaugino condensation spontaneously breaks $\Z_{2M}\to \Z_2$, so there are $M$ gauge theory vacua. As we transform $c\to c+2\pi$, we move from one gauge theory vacuum to the next, and the gaugino condensate (which corresponds to the local deformation parameters on the gravity side) picks up a phase $\exp(2\pi i/M)$. This is as in Sect.~\ref{sec:LocalBackReaction} where we learned that the $M$ different vacua are reached by dialing the RR flux quanta on the $\mathcal{B}$-cycle $Q=0,\ldots,M-1$ (see \eqref{eq:FluxStabilizedValues}).

 \section{Applications}
 \label{sec:Applications}
 \subsection{Thraxions on the Quintic: Drifting Monodromy}
 \label{sec:DriftingMonodromy}
 
 In this section we will give a concrete example of a string compactification where a light thraxion appears. Along the way we identify concrete setups in which parametrically super-Planckian racetrack-type axion periodicities are possible. We choose the CY to be the quintic three-fold which is defined as the vanishing locus of a homogeneous polynomial $P$ of rank five in $\mathbb{P}^4$. Following \cite{Greene:1996cy}, it can be brought to a conifold transition point by choosing the complex structures such that 
 \begin{equation}
 P(X^0,...,X^4)=X^3 f_4(X^0,...,X^4)+X^4 g_4(X^0,...,X^4)\, ,
 \end{equation}
 where $f_4$ and $g_4$ are generic homogeneous rank four polynomials of the projective coordinates $\{X^0,...,X^4\}$ of $\mathbb{P}^4$. The conditions $P=0,\diff P=0$ are satisfied whenever
 \begin{equation}
 0=X^3=X^4=f_4(X^0,X^1,X^2,0,0)=g_4(X^0,X^1,X^2,0,0)\, .
 \end{equation}
 Since $f_4$ and $g_4$ are chosen to be generic polynomials of rank $4$, there exist $4\cdot 4=16$ distinct solutions. These are $16$ conifold points. Hence, there are $16$ vanishing three-cycles $\mathcal{A}^i$, $i=1,...,16$. Because the solution set lies on a $\mathbb{P}^2$ submanifold of $\mathbb{P}^4$, there is precisely one homology relation among them, 
 \begin{equation}
 \sum_{i=1}^{16}[\mathcal{A}^i]=0\, .
 \end{equation}
 Hence, we have a multi throat system with $n=16$ and $m=1$ so there is one light axion.
 
 Let us give two examples that differ by choices of flux numbers. In both examples we set the coefficients $A_i$ defined in \eqref{eq:thresholdcoeff2} to unity. Generically we expect these to be of order one. Inserting $\mathcal{O}(1)$ factors below does not change the physical outcome.\footnote{If some coefficients can be tuned parametrically smaller than others, new qualitative features might arise. We leave an investigation of this possibility to future research.} 
 
 \textbf{Example 1: A simple thraxion potential} 
 \begin{equation}
 M_i=(-1)^{i+1}M\, ,\quad  K^i=(-1)^{i+1}K\, ,
 \end{equation} 
 with $K/g_sM\gg 1$. Then we have $\epsilon_i\equiv (-1)^{i+1}\epsilon$, and
 \begin{align}
 W_\text{eff}(\mathcal{G})&= -16i \epsilon \sin 2\mathcal{G}/M +\hat{W}_0=16i\epsilon (1-\cos 2\mathcal{G}'/M)+W_0\, ,
 \end{align}
 with $W_0=\hat{W}_0-16\epsilon$, $\mathcal{G}'=\mathcal{G}-\pi M/4$, and small $\abs{\epsilon}\propto \exp (-2\pi K/g_sM)$. Up to the numerical pre-factor this is exactly what we found for $n=2$ and $m=1$.
 
 \textbf{Example 2: Drifting Monodromy}
 
 We now slightly detune the $F_3$ fluxes from one another:
 \begin{equation}
 \begin{split}
 M_1=M\, ,\quad 
 M_2=M+1\, ,\quad 
 M_3=-M\, ,\quad 
 M_4=-(M+1)\, ,
 \end{split}
 \end{equation}
 and $M_{i+4}=M_i$, with $K^{i}\equiv \text{sign}(M_i) |K|$, and again $K/g_s M\gg 1$. In this case
 \begin{align}
 W_\text{eff}(\mathcal{G})&\approx-8iz_0(M\sin(\mathcal{G}/M)+(M+1)\sin(\mathcal{G}/(M+1)))+\hat{W}_0\, ,
 \label{eq:DriftingMonodromy}
 \end{align}
 with $z_0\sim \exp 2\pi i K\tau/M$. Additionaly to the previous simplification, we have also neglected order one prefactors that arise from the fact that the ratios $K^i/M_i$ are not all exactly equal. Again, this is of no consequences for our purposes.
 
 The superpotential \eqref{eq:DriftingMonodromy} is a \textit{racetrack}-type superpotential\footnote{For a discussion of racetrack superpotentials in connection with the WGC, see \cite{Moritz:2018sui}.} for $\mathcal{G}$. The axion periodicity is now given by $2\pi M(M+1)$. Crucially, this implies another $M$-fold extension of the axion field range \textit{on top} of the one already discussed in the simpler examples of the double throat and the first example of this section. Clearly, one may take this even further to periodicities such as $2\pi M\cdots (M+3)$.\footnote{But not further because we have to respect the orientifold action.} Since we still only have to fulfill the requirement that the throats fit into the bulk CY, this implies the existence of a simple, concrete and explicit mechanism in string theory that can generate huge super-Planckian axion periodicities. In general the full periodicity of the superpotential is given by the least common multiple of the different RR flux numbers $M_i$. We dub this mechanism of generating a parametrically large axion monodromy \textit{drifting monodromies} since it relies on a frequency drift within a set of several finite-order monodromy effects. This is related but different from the winding idea, where a constraint forces the effective axion on a long trajectory in a multi-axion moduli space \cite{Kim:2004rp,Berg:2009tg,Hebecker:2015rya,Choi:2015fiu,Kaplan:2015fuy}. Here, by contrast, one may think of a single fundamental axion extended by several small, finite-order monodromy effects. The result of this can still be large as explained above. The intended outcome, namely to realize an effective large-$f$ axion accepting a short-wavelength oscillatory potential, is of course the same (see in particular the very recent analysis of \cite{Hebecker:2018fln}).

\begin{figure}[t]
 	\centering
 	\includegraphics[width=6cm,keepaspectratio]{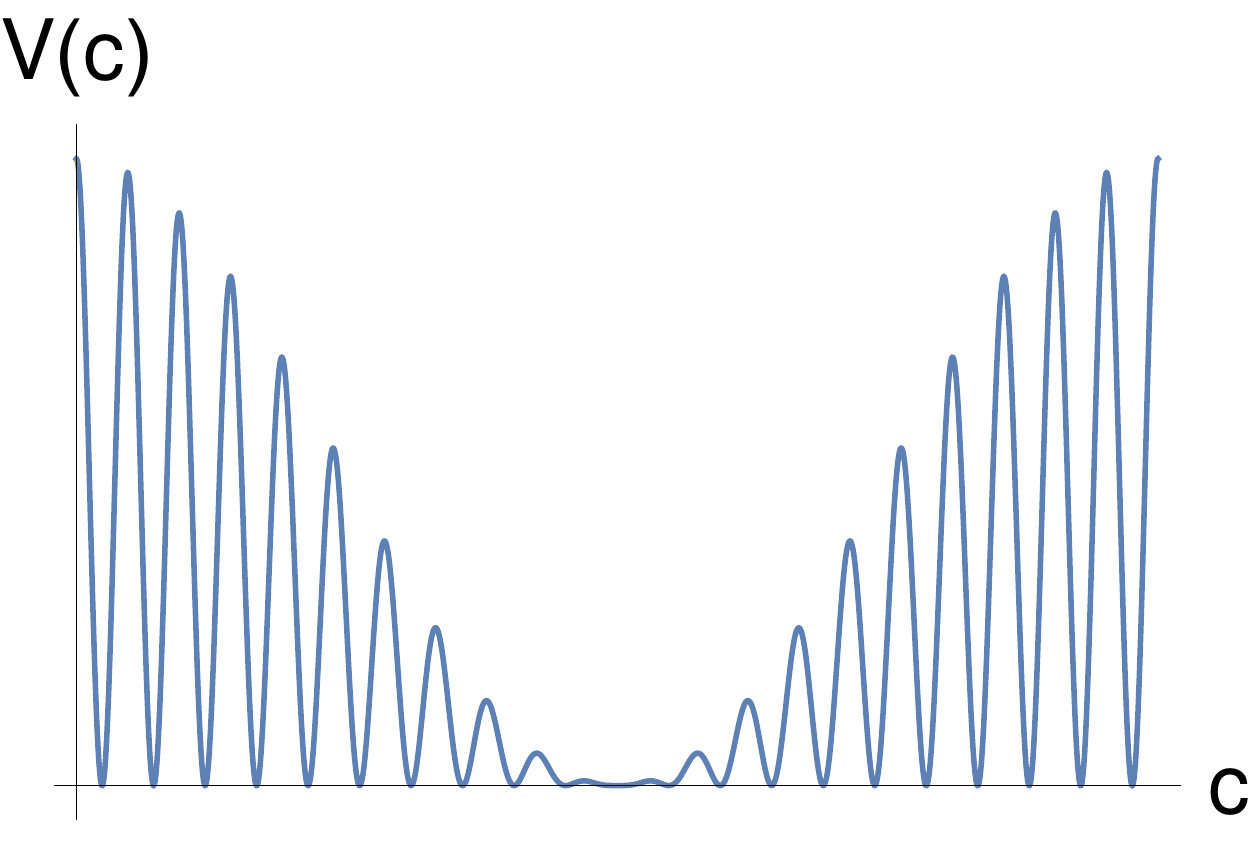}
 	\caption{The axion potential of example $2$ for the case $M=10$. There are sub-Planckian oscillations within a long super-Planckian envelope.}
 	\label{fig:cascade}
 \end{figure}

 The minima of the potential $V\propto |\del_{\mathcal{G}}W|^2$ are located along the slice $\text{Im}\,\mathcal{G}=0$ where it takes the form
 \begin{equation}
 \begin{split}\label{eq:vcDriftingMonodromy}
 V(c)&\propto \left[\cos (c/M)+\cos (c/(M+1))\right]^2\, ,\quad c\equiv \text{Re}\,\mathcal{G}\, ,\\
 &\propto \cos^2 \left(\frac{2M+1}{2M(M+1)}c\right)\cdot\cos^2 \left(\frac{1}{2M(M+1)}c\right)\,,
 \end{split}
 \end{equation}
 which has $2M+1$ distinct Minkowski vacua (see Fig.~\ref{fig:cascade}).
 
 We note that despite the long $2\pi M (M+1)$ periodicity the scalar potential oscillates on shorter wavelengths of order $2\pi M$.
This is essentially due to the rank condition \eqref{eq:rank-condition} which forces us to introduce flux numbers of both signs.\footnote{This condition is \textit{global} in the sense that it need not hold in a non-compact CY where gravity is decoupled.} We have not shown in general that suppressing such shorter wavelength oscillations in order to produce a smooth super-Planckian axion potential is impossible. At this point we only note that the condition \eqref{eq:rank-condition} presents a severe obstacle towards this.
 
 These examples also serve to illustrate that by scanning over flux numbers one may obtain a vast number of possible effective superpotentials and axionic potentials.

 \subsection{A Clash with the Weak Gravity Conjecture}
  
  In this section we would like to point out that the axion potential we have derived clashes with what has been put forward as the \textit{weak gravity conjecture for axions} \cite{ArkaniHamed:2006dz,Brown:2015iha,Heidenreich:2015wga}. We have computed the axion potential via a classical supergravity calculation. However, one may equally well associate it to non-perturbative effects in the KS gauge theory (namely gaugino condensation), as argued in Sect.~\ref{sec:gaugegravity}. As such, (if true) the weak gravity conjecture should apply to our construction.
  
  In its form adapted to axions and instantons the conjecture states that there should exist an instanton with Euclidean action $S$ and axionic charge $q$ such that\footnote{Note that $M_\Pl/f_\text{eff}$ is the analogue of the gauge coupling for zero-forms (i.e.~axions).}
  \begin{equation}\label{eq:axwgc}
  S \leq \mathcal{O}(1)\, q\, M_\Pl/f_\text{eff}\, .
  \end{equation}
  Such instantons (if they contribute to the superpotential) generically induce terms in the scalar potential of the form
  \begin{equation}
   V(\phi)\supset e^{-S}(1-\cos(q\,\phi/f_\text{eff}))\, ,
  \end{equation}
  where $\phi$ is the canonically normalized axion. Thus, $2\pi f_\text{eff}/q$ is the canonically normalized periodicity of the term in the (super)potential that is generated by a given instanton.
  
  By comparison with the above we may associate an (effective) Euclidean instanton action to each of the leading exponentially suppressed terms in the axion (super)potential\footnote{Taking the correspondence with instantons seriously, these are ($\frac{1}{2}$BPS-)instantons.}.
  \begin{equation}
   S_\text{eff}^i\approx 3\log(1/w_\IR^i)\approx 2\pi \frac{K^i}{g_s M_i}\, .
  \end{equation}
  As computed in App.~\ref{Appendix:AxionDecayConstant}, in the regime where the throats marginally fit into the bulk CY, the periodicities $f_\text{eff}/q^i$ of the dominant terms in the superpotential associated to each throat $i=1,...,n$ read
  \begin{equation}
   f_\text{eff}/q_i\approx \frac{2}{3}(r_\text{D3}^i)^{1/2}\log(1/w_\IR^i)^{-1/2} M_\Pl\approx \frac{2}{3}(r_\text{D3}^i)^{1/2}\left(\frac{2\pi}{3}\frac{K^i}{g_sM_i}\right)^{-1/2}M_\Pl\, ,
  \end{equation}
  where $r_\text{D3}^i=M_iK^i/N_\text{D3}^\text{flux}$ is the fraction of the total D$3$ brane charge of three-form fluxes which is stored in the $i$-th throat.
  Hence,
  \begin{equation}
   S_\text{eff}^i\cdot f_\text{eff}/q^i\sim 2 (r_\text{D3}^i)^{1/2} \sqrt{\log (1/w_\IR^i)}\,  M_\Pl\approx 2(r_\text{D3}^i)^{1/2}\sqrt{\frac{2\pi}{3}\frac{K^i}{g_sM_i}}M_\Pl\, .
  \end{equation}
  In the regime $w_\IR\ll 1$ (i.e.~$K^i\gg g_sM_i$) the r.h.~side is parametrically larger than $\mathcal{O}(1)$ so the objects that generate the relevant terms in the superpotential do not satisfy a weak gravity conjecture bound.
  
  Of course as is always true in string theory compactifications \cite{Banks:2003sx} there \textit{does} exist a tower of instantons that satisfies the weak gravity bound \eqref{eq:axwgc} but generates no monodromy.\footnote{In our case, these are Euclidean D$1$ strings wrapping representative $S^2$'s in the UV, compare Sect.~\ref{sec:Uplifting}.} It is also apparent that these instantons occupy a sub-lattice of the full charge lattice.
  This sub-lattice corresponds to all the possible wrapping numbers of a Euclidean D$1$ string. However, in our setup this sub-lattice can be made parametrically \textit{coarse}.\footnote{Hence we seem to realize explicitly the loop hole mentioned in footnote 25 of ref. \cite{Heidenreich:2016aqi}.} Let us illustrate this with a concrete example: We consider a variant of the \textit{drifting monodromies} example given in Sect.~\ref{sec:DriftingMonodromy}, with flux numbers $M_i\in\{5,6,7,8,-5,-6,-7,-8\}$. The axion decay constant is enhanced by the least common multiple of $5,6,7,8$ which is $840$. The instantons that satisfy \eqref{eq:axwgc} respect the periodicity of the axion \textit{before} monodromy. Thus the possible charges take values in $840\mathbb{Z}\subset \mathbb{Z}$. Clearly, a lattice WGC is parametrically violated, while a sub-lattice WGC \cite{Heidenreich:2015nta,Heidenreich:2016aqi} (see also \cite{Andriolo:2018lvp}) is always satisfied but with parametrically coarse sub-lattice. Note that generically these instantons only give rise to sub-leading corrections to the scalar potential (if they contribute at all), compare Sect.~\ref{sec:Uplifting}.
  
  However, we observe that the $n$ effective instantons that \textit{do} give the dominant contribution to the superpotential satisfy a relation
  \begin{equation}\label{eq:axwgctilde}
  S_{\text{eff}}^i\leq \frac{4}{3}r_\text{D3}^i(q^i\, M_\Pl/f_\text{eff})^2\, ,\quad \forall i=1,...,n\, .
  \end{equation}
  Together with $S_\text{eff}^i>1$, as is required for controlled expansion in powers of $e^{-S_\text{eff}^i}$, this \textit{still} implies the existence of short wavelength harmonics in the superpotential. This motivates a closer look at the full spectrum of our effective instantons to which we now turn.

  \subsection{The Spectrum of Effective Instantons}
  We now set aside the spectrum of instantons that satisfy the WGC but instead ask what are the properties of the effective instantons that generate the superpotential. We have written down the most dominant contributions to the superpotential \eqref{eq:EffSuperPotMulti} and noted that they satisfy \eqref{eq:axwgctilde}. This condition is weaker than \eqref{eq:axwgc} because at fixed control parameter $S_\text{eff}^i\overset{!}{>}1$ the value of $qM_\Pl/f_\text{eff}$ is constrained to be bigger only than $\sqrt{S_\text{eff}}$ rather than $S_\text{eff}$. Nevertheless, as a consequence the dominant effective instantons give rise to sub-Planckian wavelength oscillations in the scalar potential.
  
  In general it is easy to see that beyond these most dominant effective instantons there exists a whole $\Z^{n}$ lattice worth of these effective instantons\footnote{We consider general combinations of holomorphic and anti-holomorphic instantons.}. These correspond to the higher orders in the $|z_i|$, $i=1,...,n$, that we have neglected in Sect.~\ref{sec:4dSugra}. Therefore, the $n$ dominant effective instantons serve as $n$ basis vectors of the lattice $\Z^n$ and a general effective instanton is labeled by an effective charge vector $\vec{k}\in \Z^n$. The bound \eqref{eq:axwgctilde} for these general effective instantons reads
  \begin{equation}
  S_\text{eff}^{\vec{k}} ~\leq~ \frac{4}{3}M_\Pl^2\sum_{i=1}^{n} |k^i| \,r_\text{D3}^i\, \left(\frac{q^i}{f_\text{eff}}\right)^2\, ,
  \end{equation}
  and the r.h.~side defines a ($1$-)norm on $\Z^n$.
  
  The effective instantons can also be embedded into the full one-dimensional charge lattice $\Z$ of the preceding section via $\vec{k}\longmapsto \sum_{i=1}^{n}k^i q^i$, although they do not satisfy the WGC bound \eqref{eq:axwgc}. 
  
  Again, this is perhaps best understood using the explicit example given in the preceding section. The dominant instantons have charges
  \begin{equation}
  q^i\in \left\{\pm\frac{840}{M_i}\right\}_i= \left\{\pm\frac{840}{8},\pm\frac{840}{7},\pm\frac{840}{6},\pm\frac{840}{5}\right\}=\{\pm 105,\pm 120,\pm 140,\pm 168\}\, .
  \end{equation}
  They induce harmonics in the superpotential with periodicities $f_\text{eff}/q^i$. Due to the relation \eqref{eq:axwgctilde} precisely these combinations are restricted to be sub-Planckian as long as $S_\text{eff}^i>1$. They can be understood to occupy a coarse $105\Z+120\Z+140\Z+168\Z$ sub-lattice of $\Z$ while the Euclidean D1-instantons satisfying the WGC occupy an even coarser sub-lattice $840\Z\subset \Z$.
 \subsection{Axion Phenomenology}
  
  We have identified a string theory axion with remarkable properties. It is parametrically lighter than the tower of states that is usually associated to strongly warped regions $m_\text{tower} \propto w_\IR M_\Pl$. The axion mass can be tuned almost independently of the periodicities of the dominant oscillations in the scalar potential, since we have $m \propto w_\IR^3 M_\Pl$, while the oscillation period $f_\text{eff}/q$ of the scalar potential depends only weakly on the warp factor $f_\text{eff}/q \sim M_\Pl/\sqrt{\log(w_\IR^{-1})}$. Conversely, the mass scales unusually strongly with the oscillation wavelength, 
    \begin{equation}
    \frac{m^2}{M_\Pl^2}\propto w_\IR^6 \approx e^{-2S_{\text{eff}}}\approx \exp\left(-\alpha\left(\frac{qM_\Pl}{f_\text{eff}}\right)^2\right)\, ,
    \label{eq:MassScaling}
    \end{equation}
    with $\alpha=\calO(1)$.
    
  In contrast most other stringy axions usually satisfy the relation \cite{Arvanitaki:2009fg}
  \begin{equation}
  \frac{m^2}{M_\Pl^2}\sim \exp\left({-\alpha \frac{qM_\Pl}{f_\text{eff}}}\right)\, ,
  \end{equation}
  As such the thraxion assumes a rather special place in the string theory landscape. This is potentially interesting for axion phenomenology. We refer the reader to \cite{Arvanitaki:2009fg} for a range of phenomenological applications for different axion mass scales.
  
  We have to emphasize that at least in the simplest setups our axion is not a generic inflaton candidate because of the generic presence of dominant sub-Planckian wavelength modes in the scalar potential, \textit{despite} the large monodromy enhancement of the effective axion decay constant.

  \subsection{Uplifting}
  \label{sec:Uplifting}
  
  We would like to briefly comment on some possible scenarios of uplifting. To actually implement these ideas in concrete models involves the complicated interplay of different effects. We highlight the assumptions needed for the following two scenarios in turn. Both of them are based on the idea of adding an oscillating potential of different wavelength to the known thraxion potentials of the form \eqref{eq:vc} or \eqref{eq:vcDriftingMonodromy}.
  
  Uplifting requires as a precondition, that a full mechanism of K\"ahler moduli stabilization is in place. Stabilizing the K\"ahler moduli by definition breaks the no-scale property of GKP-type flux compactifications. Hence, in a full setup of our multi throat system there are two sources of no-scale breaking -- the Calabi-Yau breaking potential of our $C_2$-axion(s), and also the scalar potential that stabilizes K\"ahler moduli. In general we expect these two sources of no-scale breaking to mix non-trivially, and we leave a detailed analysis of this for future research. For the rest of this discussion we now assume that these subtleties get resolved for both KKLT and large volume scenario (LVS)-type setups of K\"ahler moduli stabilization.

  We now wish to look at situations where $c$-dependent corrections to the K\"ahler potential may become relevant. This is certainly the case in the regime $\abs{W_0} \sim 1$, leading us to consider LVS-like moduli stabilization \cite{Balasubramanian:2005zx}. Potentially interesting corrections may arise from Euclidean D1-brane instantons that wrap members of the family of two-spheres that vanish at the tips of the conifold. Since the cycle is trivial in homology we expect no corrections to the superpotential but at most corrections to the K\"ahler potential of the form
  \begin{align}
  \delta e^{-K_1} &\sim   \mathcal{C}e^{-S_\text{DBI}-iS_\text{CS}}+c.c.\, ,
  \label{eq:correction}
  \end{align}
  with $\mathcal{C} = \calO(1)$ and DBI and CS actions
  \begin{equation}
  S_\text{DBI}= \frac{1}{g_s}\frac{\text{Vol}(S^2)|_{\UV}}{2 \pi \ap}\, , \quad \text{and}\quad  S_{\text{CS}}=\frac{1}{2 \pi \ap}\int_{S^2}\left(\sum_{p}C_p\right)\wedge e^{B}|_{2-\text{form}}=\text{Re}\,\mathcal{G}\, .
  \end{equation}
  Here, we have evaluated the DBI action on a representative sphere in the UV, i.e.~in the bulk CY. This is because we expect such a representative to give the dominant contribution: As explained in App.~\ref{sec:AxionPeriodicity}, there are different two-spheres at a given radial coordinate in the throat that are labeled by a $U(1)$ phase and that all share the same volume. As we scan over this phase, the corresponding integrals of $C_2$ at a given radial coordinate pass through their fundamental domain. Therefore, integrating over all Euclidean brane instantons on the two-spheres should cancel all contributions due to the oscillatory behavior of the correction \eqref{eq:correction}. This is consistent with the fact that after accounting for backreaction of the phases of the throat deformations the $C_2$ field excursion cannot be measured in the local throats. In the analysis of Sect.~\ref{sec:4dSugra}, we extended this result to $\text{Re}(\mathcal{G})$, i.e.~to $C_2 - C_0 \, B_2$. In passing towards the UV, our description of the throat breaks down. In particular, we do not expect the different sphere representatives to all share the same volume. Thus, we expect non-vanishing instanton corrections.
  
  Using $\text{Vol}(S^2)|_{\UV} \gtrsim R_\text{throat}^2 \propto (g_s M K)^{1/2} \ap$, this leads to corrections to the scalar potential of the form 
  \begin{equation}
  \delta V \lesssim e^{-\alpha \sqrt{\frac{KM}{g_s}}} (1-\cos(\text{Re}\,\mathcal{G}))\, ,
  \end{equation}
  with $\alpha = \calO(1)$. Assuming that the exponentially small prefactors of the classical warping suppressed potential \eqref{eq:vc} (or that of example $1$ of Sect.~\ref{sec:DriftingMonodromy}) and the non-perturbative correction terms are of the same order, it is feasible that additional local minima appear in the scalar potential that could in principle lift to meta-stable non-supersymmetric minima, possibly even de Sitter vacua. 
  
  The exponential terms are of comparable magnitude when
  \begin{equation}
  \sqrt{\frac{K}{g_sM}} \gtrsim M \,.
  \end{equation}
  In F-theory models with large Euler characteristic we do not see an immediate obstacle to realizing this.
  
\begin{figure}[t]
 	\centering
 	\includegraphics[width=6cm,keepaspectratio]{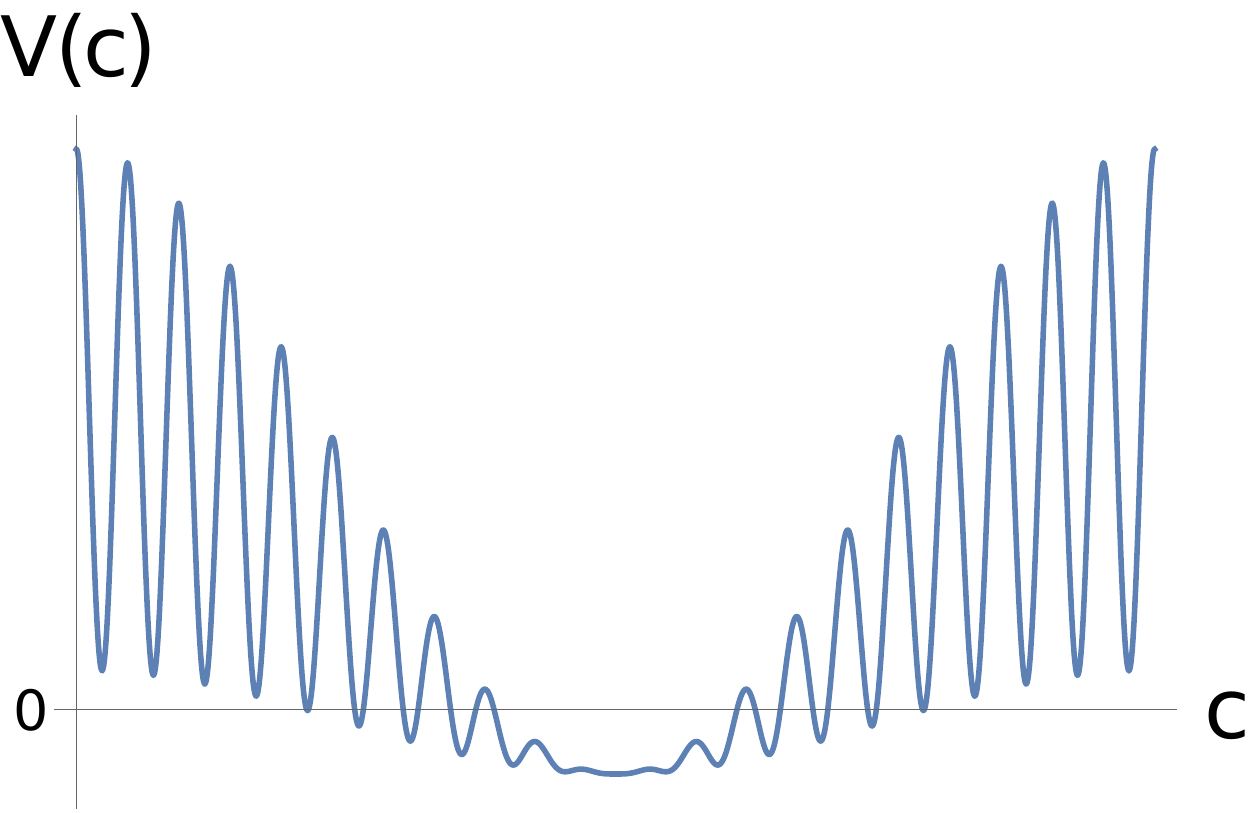}
 	\caption{The axion potential of Fig.~\ref{fig:cascade} with an additive correction $\delta V(c) \propto \text{const.} +  \cos(c/f_\text{eff})$ that shares the periodicity of the superpotential.}
 	\label{fig:uplift}
  \end{figure}

  We may turn this around and add large-wavelength corrections to shorter-wavelength oscillations such as those of the example of \textit{drifting monodromies} given in Sect.~\ref{sec:DriftingMonodromy}. On the large scale of $f_\text{eff} \sim M M_\Pl$ there are several Minkowski vacua of the potential \eqref{eq:vcDriftingMonodromy}, compare Fig.~\ref{fig:cascade}. It is conceivable that these are uplifted to de Sitter vacua once further corrections to the potential are taken into account. This might happen automatically when the no-scale properties of the K\"ahler potential \eqref{eq:Kahlerpot} get broken by perturbative or non-perturbative corrections, since we know that the existence of Minkowski vacua strongly depends on the cancellation of different terms in the scalar potential. We expect these scalar potential corrections to follow the periodicity of the superpotential, which is given by the super-Planckian decay constant $f_\text{eff}$. An optimistic sketch of this is illustrated in Fig.~\ref{fig:uplift}. A different approach might be to consider \textit{drifting monodromies} in which we allow for hierarchies between fluxes $M_i$ of individual throats.
  
  We leave a more thorough investigation of these uplifting ideas for future research.
  
 \section{Conclusion and Discussion}
 \label{sec:Conclusion}
 
 In this paper we have shown that a novel type of axion-like particle is present in many flux compactifications of type IIB string theory. While its exponentially small mass is due to the existence of strongly red-shifted regions in the compactification manifold (warped throats), it is parametrically lighter than the red-shifted tower of KK-modes that is usually associated with such throats. We would like to emphasize that for fixed value of the axion decay constant the \textit{thraxion} mass is far smaller than all other stringy axions that we are aware of. Moreover, we are able to find explicit models with even parametrically super-Planckian axion decay constants (but with generically dominant sub-Planckian oscillations in the scalar potential). As such the thraxion assumes a rather special place in the string theory landscape that is potentially interesting for axion phenomenology.
 
 The existence of this type of ultra-light axion is intimately linked to so-called conifold transitions between (topologically) distinct CY manifolds. Light thraxions arise whenever fluxes drive a CY close to the transition point. Moreover they come paired with light scalar degrees of freedom (`saxions') which are the relevant light degrees of freedom that control the global stabilization of the multi throat system that arises near such a point in moduli space. At the perturbative level the saxion is as light as the axion while non-perturbative K\"ahler moduli stabilization effects would generically lift this degeneracy. The extremely low (s)axion mass implies that multi throat systems are surprisingly weakly stabilized.
 
 We now summarize the key steps in the derivation of the (s)axion potential. Throughout most of this paper we have focused on the case of a double throat system. As shown in \cite{Hebecker:2015tzo} in such a setting there exists a light axion mode that can be thought of as the integral of the RR two-form $C_2$ over a family of spheres that degenerate at the infrared ends of the two throats. When holding the geometry fixed, a finite field excursion leads to the formation of a flux/anti-flux pair at the two ends, so an appropriately red-shifted potential energy $V\sim w_{\IR}^4$ is induced where $w_{\IR}$ is the infrared warp factor of the throats. One of the key observations made in this paper is that when the throats are long there exist \textit{two} light complex scalars $z_1,z_2$ that control the infrared geometry of the individual throats. This is despite the fact that only a diagonal combination of the two is an actual complex structure modulus of the CY. A finite field excursion of the axion mode drives $z_1$ and $z_2$ away from complex structure moduli space, i.e.~to $z_1\neq z_2$. After this geometrical backreaction is accounted for, locally in each throat supersymmetry is almost perfectly restored. We have determined the scale of the remaining scalar potential by finding the higher dimensional field profile that interpolates the 4d modes $z_1$ and $z_2$ between the two throats. It turns out to be dominated by tiny contributions from the bulk CY where the mismatch between the two throats is detected. The final potential energy scales as $w_{\IR}^6$ which is parametrically smaller than the estimate of \cite{Hebecker:2015tzo}. Since the axion kinetic term is dominated by contributions from the bulk geometry, the axion mass is of order $w_{\IR}^3$ which is parametrically smaller than the infrared mass-scale of the local throats. Furthermore, the final axion potential turns out to be periodic, and the periodicity is enhanced by a flux number $M$ via monodromy. In other words, the available amount of axion monodromy after backreaction is finite.\footnote{We note that the local two-node setup of~\cite{Aganagic:2006ex} on the deformed side of the geometric transition resembles our setup in the non-compact flux-less limit. Hence, it is tempting to speculate that fluxed (warped) versions of this, as proposed in \cite{McAllister:2008hb} and further explicated in \cite{Retolaza:2015sta} with its D$5$-$\overline{\text{D}5}$ (or NS$5$-$\overline{\text{NS}5}$) system, will show similar backreaction effects as discussed in our paper, which might significantly affect the axion monodromy potential. It is an open challenge to construct the geometry for the wrapped $\text{NS}5-\overline{\text{NS}5}$-pair at a level such that this question can be addressed explicitly.} The low scale of the axion potential implies that throughout the compact axion field space the local throats are essentially frozen supersymmetrically.
 
 We have cross-checked and expanded on the 10d double throat calculation in several ways: First, we have shown that a natural proposal for the extension of the GVW flux superpotential leads to an axion potential that matches precisely our 10d conclusions. Moreover, in this framework it is immediate that the axion is paired with a saxion that can be thought of as the integral of the NS two-form $B_2$ over the same family of spheres. Together they form the complex scalar component of a chiral multiplet. Furthermore we have extended the discussion to the case of a general multi throat system where $n$ three-cycles degenerate near a conifold transition point, with $m$ homology relations among them. There are $n$ `local complex structure moduli' $z_i$ that are relevant for the discussion of which only $n-m$ linear combinations correspond to complex structure moduli space. In this setup there are $m$ families of two-spheres that can be used to define $m$ (s)axions. Again, a finite field excursion of the (s)axions drives the local deformation parameters $z_i$ away from complex structure moduli space and a small periodic potential remains which is of order $w_\IR^6$. We have checked our conclusions also from the perspective of the Klebanov-Strassler gauge theory \cite{Klebanov:2000nc,Klebanov:2000hb} that is believed to be the gauge theory dual of the throat solution. The (s)axions set a combination of gauge couplings at the UV-ends of the throats and receive a non-perturbative superpotential from gaugino condensation in the infrared. This superpotential again matches precisely the one we have obtained from \textit{classical} 10d/4d supergravity. As is common in the gauge/gravity duality, a classical effect on one side matches a non-perturbative quantum effect on the other.
 
 Our construction can be used to investigate a vast number of axion-models, by scanning over different conifold transition points of CY threefolds and three-form flux quanta. We have illustrated this by giving two examples based on a well-studied conifold transition point of the quintic threefold. With one of these examples we exhibit a mechanism that may be able to generate parametrically super-Planckian axion periodicities. This happens when several throats carry different flux numbers $M_i$ that each give rise to a finite monodromy enhancement. The overall enhancement is given by the least common multiple of all the flux numbers. In the regime considered, the validity of the effective field theory is \textit{not} undermined by the appearance of a large number of light states below or near the axion mass-scale as is usually expected for large $f$ \cite{Banks:2003sx,Heidenreich:2015nta,Heidenreich:2016aqi} or as parametrically large geodesics in field space are traversed \cite{Ooguri:2006in,Klaewer:2016kiy,Blumenhagen:2017cxt,Grimm:2018ohb,Heidenreich:2018kpg,Lee:2018urn,Lee:2018spm,Grimm:2018cpv}. We call this mechanism to generate large axion periodicities \textit{drifting monodromies}. Nevertheless, we identify a global constraint among the flux numbers that presents a serious obstacle to actually realizing a scalar potential with parametrically sub-dominant sub-Planckian oscillations.
 
 Clearly, the examples we have given form only a tiny subset of possible \textit{thraxion} models. Moreover, these models are a promising playground for testing swampland conjectures such as the weak gravity conjecture for axions. The simplest models already suggest that the simplest form of the conjecture need not hold in general. We have briefly commented on thraxion-related ideas for uplifting to de Sitter minima. However, whether or not these ideas can be realized in a controlled way is left for future work. This would require a detailed understanding of the interplay between the thraxions and no-scale breaking effects such as gaugino condensation on seven-branes \cite{Kachru:2003aw} or perturbative $\alpha'$ corrections~\cite{Becker:2002nn}, which are needed for full moduli stabilization \cite{Kachru:2003aw,Balasubramanian:2005zx}.
 
 \vspace{1cm}
 {\large \textbf{Acknowledgements}}
 We would like to thank Federico Carta, Richard Eager, Markus Dierigl, Thomas Grimm, Lukas Hahn, Craig Lawrie, Evan McDonough and Ander Retolaza for useful discussions and comments. This work is supported by the ERC Consolidator Grant STRINGFLATION under the HORIZON 2020 grant agreement no. 647995.

 \appendix
  
 \section{Deformed Conifold for General Complex Structure Modulus}
 \label{sec:Notation}
 
  The unwarped internal geometry before backreaction by fluxes is that of the deformed conifold \cite{Candelas:1989js}. It can be embedded in $\C^4$ via
  \begin{equation}
   \sum_i w_i^2 = z \,,
   \label{eq:Embedding}
  \end{equation}
  where $w_i$ are complex coordinates and $z$ is some complex parameter. For now, we set $z = \abs{z} \in \R^+$. One can check, that the five-dimensional hypersurface at a given fixed radial coordinate $\rho^2 \equiv \sum \abs{w_i}^2 > \abs{z}$ allows for a transitive $SU(2)\times SU(2)$ action\footnote{Acting as $SO(4)$ on the vector $(w_1,w_2,w_3,w_4)^T$ in the obvious way \cite{Herzog:2001xk}. It is more convenient to parametrize the embedding via the matrix $T = \frac{1}{\sqrt{2}} \sum w_i \sigma_i$, with Pauli matrices $\sigma_i$. General solutions to $\det T = - \frac{\abs{z}}{2}$ and $\rho^2 = \Tr{T^\dagger T}$ then take the form $T = L T_0 R$, with a specific solution $T_0(\abs{z},\rho)$ and $(L,R) \in SU(2) \times SU(2)$. The $SU(2)\times SU(2)$ action is now the obvious one $(L,R) \to (U_L L, U_R R)$ \cite{Minasian:1999tt}.} with isotropy group $U(1) \subset SU(2)\times SU(2)$. The space can therefore be written as the homogeneous space $\frac{SU(2)\times SU(2)}{U(1)}$. At $\rho^2 = \abs{z}$, the isotropy group is enhanced to a diagonal $SU(2) \subset SU(2) \times SU(2)$, therefore leading to the homogeneous space $\frac{SU(2)\times SU(2)}{SU(2)} \sim \frac{SO(4)}{SO(3)} = S^3$. Given that topologically $\frac{SU(2)\times SU(2)}{U(1)} = S^3 \times S^2$, we arrive at the picture of a family of two-spheres along some radial coordinate fibered over a three-sphere, where the two-sphere vanishes or collapses at the apex, while the three-sphere stays finite everywhere.
  
  Following ref. \cite{Minasian:1999tt}, we may derive the most general $SU(2)\times SU(2)$-invariant metric, that is also invariant under a $\Z_2$ exchange of the two $SU(2)$-factors. It takes the form
  \begin{equation}
  \begin{split}
   \diff s^2 =& \left(B^2 + H^2 \right) \left(\left(g^1\right)^2 + \left(g^2\right)^2\right) + C^2 \left(\left(g^3\right)^2 + \left(g^4\right)^2\right) \\
   &+D^2 \left(\diff \rho^2 + \left(g^5\right)^2\right) + 2 C H \left(g^1 g^4 - g^2 g^3\right)\,.
   \label{eq:MetricGeneralCoefficients}
  \end{split}
  \end{equation}
  Here, we used the vielbein
  \begin{equation}
   g^1 = \frac{e^1-e^3}{\sqrt{2}}\,, \quad g^2 = \frac{e^2-e^4}{\sqrt{2}}\,, \quad g^3 = \frac{e^1+e^3}{\sqrt{2}}\,, \quad g^4 = \frac{e^2+e^4}{\sqrt{2}}\,, \quad g^5 = e^5 \,,
   \label{eq:Vielbein}
  \end{equation}
  with
  \begin{equation}
  \begin{split}
   e^1 = -\sin \theta_1 \diff \phi_1\,, \quad  e^2 = \diff \theta_1\,,& \quad  e^3 = \cos \psi \sin \theta_2 \diff \phi_2-\sin \psi \diff \theta_2\,,\\
   e^4 = \sin \psi \sin \theta_2 \diff \phi_2 + \cos \psi \diff \theta_2\,,& \quad e^5 = \diff \psi + \cos \theta_1 \diff \phi_1 + \cos \theta_2 \diff \phi_2\,.
  \end{split}
  \end{equation}
  
  The real coefficients $B$, $C$, $D$ and $H$ are functions of $\rho$ and $\abs{z}$. We impose the constraints on the coefficients from  the embedding \eqref{eq:Embedding} and insert the definition of the radial coordinate, as well as demand Ricci-flatness of the metric \eqref{eq:MetricGeneralCoefficients}. The resulting CY-metric with $SU(2)\times SU(2)$-isometry is now best expressed in the coordinate $\tau\geq 0$ defined via $\rho^2 = \abs{z} \cosh \tau$ \cite{Minasian:1999tt}, 
  \begin{equation}
  \begin{split}
   \diff s_\text{dc}^2 = D_\text{dc}^2(\tau) \left(\diff\tau^2 + \left(g^5\right)^2\right) +  B_\text{dc}^2(\tau) \left(\left(g^1\right)^2 + \left(g^2\right)^2\right) + C_\text{dc}^2(\tau) \left(\left(g^3\right)^2 + \left(g^4\right)^2 \right) \,,
  \end{split}
  \end{equation}
  with
  \begin{align}
    B^2_\text{dc}(\tau) &= \abs{z}^\frac{2}{3} \frac{K(\tau)}{2} \sinh^2\left(\tau/2\right)\,, \quad 
    C^2_\text{dc}(\tau) = \abs{z}^\frac{2}{3} \frac{K(\tau)}{2} \cosh^2\left(\tau/2\right)\,, \\
    D^2_\text{dc}(\tau) &= \abs{z}^\frac{2}{3} \frac{1}{6 K(\tau)^2}\,, \quad \text{where}\quad 
    K(\tau) \equiv \frac{\left(\sinh\left(2 \tau\right) - 2 \tau\right)^{\frac{1}{3}}}{2^{\frac{1}{3}} \sinh(\tau)} \,,
  \end{align}
  and $H_\text{dc} \equiv 0$.
  
  We now generalize this metric to arbitrary values of $z = \abs{z} e^{i \varphi} \in \C$. This is most easily done by considering the action of $e^{i \varphi} \in U(1)_R$ on the coordinates
  \begin{equation}
   w_i \to w_i \, e^{i \varphi/2} \,.
   \label{eq:U1RActionCoord}
  \end{equation}
  Looking at the embedding \eqref{eq:Embedding}, this gives the rotation $\abs{z} \to \abs{z} e^{i \varphi}$ we are after. After some calculation, one can reinterpret the rotation of coordinate $w_i$ as an action on the coefficients $B$, $C$, $D$ and $H$. The new coefficients after applying a $U(1)_R$ rotation to second order in $\varphi$ read
  \begin{equation}
  \begin{split}
   B &= B_\text{dc} + \varphi^2 \frac{B_\text{dc}}{C_\text{dc}} \frac{C_\text{dc}^2 - B_\text{dc}^2}{8 C_\text{dc}} \,,\\
   C &= C_\text{dc} + \varphi^2 \frac{B_\text{dc}^2 - C_\text{dc}^2}{8 C_\text{dc}} \,,\\
   H &= \varphi \frac{C_\text{dc}^2 - B_\text{dc}^2}{2 C_\text{dc}} \,.
   \label{eq:U1TransformedMetric}
  \end{split}
  \end{equation}
  In the full result (i.e.~beyond quadratic order in $\varphi$) the periodicity $\varphi\longrightarrow \varphi+2\pi$ is of course preserved.
  
  We now want to determine the metric at large $\rho^2\gg |z|$. For $\varphi=0$ one may express it in the form
  \begin{equation}
   \diff s^2 = f_0(r)^2 \diff r^2 + r^2 \left( \frac{1}{9} f_5(r)^2 (g^5)^2 + \frac{1}{6} \sum_{i=1}^4 f_i(r)^2 (g^i)^2 \right)\, ,
  \end{equation}
  where we have introduced the asymptotic conifold radial coordinate $r = \sqrt{\frac{3}{2}}\rho^{2/3}$. Defining $\epsilon(r) \equiv \frac{\abs{z}}{r^3}$, at order $\epsilon^0$ one has $f_i \equiv 1$ which corresponds to the metric of the singular conifold. However, also far away from the deformation $\epsilon \ll 1$, we see some remnant of the apex geometry
  \begin{equation}
  \begin{aligned}
   f_0^2(r) = f_5^2(r) &= \frac{9}{r^2} D^2\left(\tau(r)\right) = 1 + \calO(\epsilon^2) \,,\\
   f_1^2(r) = f_2^2(r) &= \frac{6}{r^2} B^2\left(\tau(r)\right) = 1 + \left(\frac{3}{2}\right)^{3/2} \epsilon + \calO(\epsilon^2) \, ,\\
   f_3^2(r) = f_4^2(r) &= \frac{6}{r^2} C^2\left(\tau(r)\right) = 1 - \left(\frac{3}{2}\right)^{3/2} \epsilon + \calO(\epsilon^2) \, .
   \label{eq:conifoldexpansion}
  \end{aligned}
  \end{equation}
  Neglecting numerical $\calO(1)$-factors, we arrive at
  \begin{equation}
   \diff s^2 = \diff r^2 + r^2 \left\{\diff \Omega_\T^2 + \epsilon(r) \left[\left(g^1\right)^2 + \left(g^2\right)^2 - \left(g^3\right)^2 - \left(g^4\right)^2 \right] \right\}+\mathcal{O}(\epsilon^2)\,.
  \end{equation}
  Using the transformation behaviour of the coefficients of the angular terms \eqref{eq:U1TransformedMetric}, it is now straightforward to write down the same asymptotic expansion for non-zero $\varphi$
  \begin{equation}
  \begin{split}
   \diff s^2 &= \diff r^2 + r^2 \left\{\diff \Omega_\T^2 + \epsilon(r) d\Omega_5^2(\varphi) \right\}\,.\\
   d\Omega_5^2(\varphi)&\equiv(1+\varphi^2)\left(\left(g^1\right)^2 + \left(g^2\right)^2-\left(g^3\right)^2 - \left(g^4\right)^2 \right)  + \varphi \left( g^1 g^4 - g^2 g^3 \right)
  \end{split}
  \end{equation}

  Finally, we consider the full 10d metric found as a solution to 10d SUGRA with $M$ units of $F_3$-flux on the three-sphere described above and some $H_3$-flux (unquantized in the non-compact setting) on the dual cycle \cite{Klebanov:2000hb}
  \begin{equation}
   \diff s^2 = w^2(r) \eta_{\mu \nu} \diff x^\mu \diff x^\nu + w^{-2}(r) \diff s_\text{dc}^2 \,,
  \end{equation}
  where the warp factor goes like
  \begin{equation}
   w(r)^2 = \frac{r^2}{g_s M \ap}\frac{1}{\sqrt{\log\left(\frac{r^3}{|z|}\right)}}\, ,
   \label{eq:WarpFactor}
  \end{equation}
  for $\frac{r^3}{\abs{z}} \gg 1$. For arbitrary $\varphi$, we will work with the asymptotic metric
  \begin{equation}
   \diff s^2 = w^2(r) \eta_{\mu \nu} \diff x^\mu \diff x^\nu + w^{-2}(r) \left[ \diff r^2 + r^2 \left( \diff \Omega_\T^2 + \epsilon(r) \diff \Omega_5(\varphi)^2 \right) \right] \,.
   \label{eq:FullMetric}
  \end{equation}

 \section{The Axion Potential in the Local Throat}
 \label{sec:AxionPeriodicity}
 
  In the main text we have repeatedly made use of the fact that the $C_2$- and $B_2$-axions $c$ and $b$ can only enter the scalar potential that is generated in the local throat in certain combinations with the `local complex structure' of the throat, namely the real and imaginary part of
  \begin{equation}\label{eq:local-physical-exc}
   M \log(z)- i \mathcal{G}\, ,
  \end{equation}
  where $\mathcal{G}=c-\tau b$, and $z$ is the `local complex structure'. Here, we would like to derive this without using `local flux stabilization' as in Sect.~\ref{sec:CYBreakingPotential} but rather rely only on asymptotic properties of the KS/KT solution \cite{Klebanov:2000nc,Klebanov:2000hb}. For simplicity we will set the RR zero form to zero, i.e.~$\tau=i g_s^{-1}$.
  
  We cut off the throat at a radial coordinate $r_{\UV}$ and define the $b$-axion at that value of the radial coordinate. Since the $B_2$ profile runs along the radial direction \cite{Klebanov:1999rd}, changing $b\longrightarrow b+\delta b$ can be realized by choosing a \textit{different} UV-cutoff $r_{\UV}'$. Since the absolute value of the complex structure is defined in units of the UV-cutoff it scales as
  \begin{equation}
   z\longrightarrow e^{-\frac{\delta b}{g_sM}}z\, .
  \end{equation}
  Since this is just a coordinate transformation from the perspective of the local KS throat, the combination $g_sM \log(|z|)+b$ cannot appear in the scalar potential that is generated within the throat. It acquires physical meaning only if the throat is cut off at fixed, finite $r_{\UV}$.
  
  Similarly, in the limit $\abs{z}/r^3 \to 0$ the RR three form takes the form
  \begin{equation}
   F_3=2 \pi \ap \, M\left(g^5+\diff c/M\right)\wedge \omega_{\Sigma}\, ,
  \end{equation}
  where $g^5=\diff \psi+...$ is given by \eqref{eq:Vielbein}, and $\omega_{\Sigma}$ is the normalized harmonic $2$-form of $T^{1,1}$. The field $c(x)$ transforms like a Goldstone boson under (local) coordinate transformations \cite{Herzog:2002ih}
  \begin{equation}
   \psi\longrightarrow \psi+2\omega(x) \, ,\quad  c(x)\longrightarrow c(x)-2M\omega(x)\, .
  \end{equation}
  Shifting along this angular direction is an isometry of the asymptotic KS solution (called $U(1)_R$). Near the IR this is not the case precisely because (by definition) the phase of the complex structure also transforms like a Goldstone boson, see App.~\ref{sec:Notation}
  \begin{equation}
   \arg z\longrightarrow \arg z-2\omega\, .
  \end{equation}
  Again, in the local throat the combination $M\arg z+c$ has no physical meaning as it is eaten via the Higgs mechanism. Only when the throat is glued into the CY space at finite radial coordinate $r_{\UV}$ does the $c$-axion gain its independent physical meaning because the $U(1)_R$ symmetry is badly broken by the CY geometry. Putting together the real and imaginary part of $\mathcal{G} = c- \tau b$, we arrive at the conclusion that only the combination \eqref{eq:local-physical-exc} is physical when considering a single throat. A second degree of freedom only becomes physical by finiteness of the throat, i.e.~by breaking the asymptotic $U(1)_R$ symmetry. This also implies that the kinetic terms for the fields $\varphi_{1,2}$ stated in \eqref{eq:eff.Lagr1} actually take the form $[\del (\varphi_{1,2}\pm c/M)]^2$ since they arise from local throat physics. We have disregarded some unimportant off-diagonal terms in the kinetic matrix.
  
  The transformation behavior of $c(x)$ under a $U(1)_R$ can also be calculated directly \cite{Leonhardt:2017}. For this, we choose some $S^2$-submanifold of the cross-section $\T$ of the deformed conifold far in the UV and apply the $U(1)_R$ action on the coordinates as in App.~\ref{sec:Notation}. One can show that the manifold $X$ defined by the $U(1)_R$-orbit of $S^2$-submanifolds is an element of $H^3(\T)$, and therefore a multiple of the $S^3$-cycle of $\T$. It turns out that the multiplicity is $2$. Denoting a member of the family of two-spheres as $S^2(\omega)$, $\omega \in U(1)_R$, we therefore find by applying Stokes' theorem
  \begin{equation}
   \int_{S^2(0)} C_2 - \int_{S^2(2 \pi)} C_2 = \int_{X} F_3 =2 \int_{S^3} F_3 = 2 (2 \pi)^2 \ap M \,.
  \end{equation}
  Using homogeneity of the flux distribution we extrapolate $\int_{S^2(0)} C_2 - \int_{S^2(\omega)} C_2 = 4 \pi \ap M \omega$. In terms of $c$ defined on (say) $S^2(0)$, we arrive at the transformation law as stated above. Interpreting $\int_{S^2} C_2$ as a generalized Wilson line, we find the usual behavior under deforming the integration path over some region with non-vanishing associated field-strength $F_3$. Under $\pi \in U(1)_R$, the family $S^2(\omega)$ sweeps out $S^3$ completely; accordingly the Wilson line picks up the flux $\int_{S^3} F_3 = (2 \pi)^2 \ap M$.

 \section{Derivation and Solution of the 5d Equations of Motion}
 \label{sec:ExactResults}
 
 \subsection{The 5d Action for \texorpdfstring{$\varphi$}{Phi}}
  
  In App.~\ref{sec:Notation} we explained how the phase $\varphi$ of the complex structure modulus $z$ enters the metric to leading order in $\abs{z}$. In this subsection we now want to explicitly derive how we arrive at the action \eqref{eq:5dActionForPhi} from this. The 10d metric \eqref{eq:FullMetric} can be decomposed into the external and radial five-dimensional metric and the internal angular metric
  \begin{equation}
  \begin{split}
   G_{MN} &= \left(w^2 \eta_{\mu\nu} \oplus w^{-2} g_{rr}\right) \oplus \left( w^{-2} r^2 g_{\Omega,mn} \right) \,, \qquad m,n = \phi_1,\phi_2,\theta_1,\theta_2,\psi \,, \\
   g_{\Omega,mn} &= g_{\T,mn} + \epsilon(r) h_{mn} \,,\quad {g_{T^{1,1}}}_{mn}dy^mdy^n\equiv d\Omega^2_{T^{1,1}}\, ,\quad  
   h_{mn} \diff y^m \diff y^n = d\Omega_5^2(\varphi)\,,  
  \end{split}
  \end{equation}
  where $\epsilon(r) = \abs{z}/r^3$ is the $U(1)_R$-breaking parameter. As explained in Sect.~\ref{sec:CYBreakingPotential}, we promote $\varphi$ to a dynamical field $\varphi = \varphi(x^\mu,r)$. While this field extends into multiple throats, we make use of its antisymmetry in order to reduce the problem to a single throat. The corresponding boundary conditions are given below.
  
  We expand the 5d action for $\varphi$ in $\epsilon$ via the expansion of the 10d Ricci scalar about the $U(1)_R$-preserving metric
  \begin{equation}
  \begin{split}
   R(G) &= R\left(G_{U(1)_R}\right) + \epsilon(r) \nabla_m \nabla_n h^{mn} - \square \left(\epsilon(r) \, g_\T^{mn} h_{mn} \right) + \calO(\epsilon^2)\,,\\
   G_{U(1)_R} &= w^2 \eta \oplus w^{-2} g_{rr} \oplus w^{-2} r^2 g_{\T} \,, \\
   G_\text{5d} &= w^2 \eta \oplus w^{-2} g_{rr} \,.
  \end{split}
  \end{equation}
  We know that for $\varphi=\text{const.}$ we simply describe a deformed conifold with general complex deformation parameter $z = \abs{z} e^{i \varphi}$, which is by construction Ricci-flat. Therefore, we have $R(G) = 0$ in every order in $\epsilon$, such that the only non-vanishing contribution can come from derivative terms of $\varphi$
  \begin{equation}
  \begin{split}
   \square \left(\epsilon(r) \, g_\T^{mn} h_{mn} \right) &= \epsilon(r) \, g_\T^{mn} \, G_\text{5d}^{ij} \, \partial_i \partial_j h_{mn}(\varphi(x^\mu,r))\\
   &= \epsilon(r) \, g_\T^{mn} \left[ \partial_\varphi h_{mn}  \,G_\text{5d}^{ij}  \,\partial_i \partial_j  \varphi + \partial_\varphi^2 h_{mn}  \,G_\text{5d}^{ij}  \,\partial_i \varphi  \,\partial_j \varphi \right]\,.
  \end{split}
  \end{equation}
  We can write
  \begin{equation}
   g_{\T ,mn} \diff x^m \diff x^n = \frac{1}{6} \left( \left(g^1\right)^2 + \left(g^2\right)^2 + \left(\left(g^3\right)^2 + \left(g^4\right)^2 \right) \right) + \frac{1}{9} \left(g_5\right)^2 \,.
  \end{equation}
  With this and looking at the perturbation $h_{mn}$ above, one finds that to order $\epsilon$, there is no contribution to the 5d action. This is due to the fact that there is no non-zero contraction $g_\T^{mn} \partial_\varphi h_{mn}$ or $g_\T^{mn} \partial_\varphi^2 h_{mn}$. There is no term $g^1 g^4 -g^2 g^3$ in $g_\T$ and the terms coming from contractions of $\left(g^1\right)^2 + \left(g^2\right)^2 $ and $\left(g^3\right)^2 + \left(g^4\right)^2 $ cancel due to the different signs between these terms in $g_\T$ and $h$\footnote{In other words: The term in $g_\T$ is symmetric under the exchange of the $SU(2)$'s defining the cone, $(\phi_1,\theta_1) \leftrightarrow (\phi_2,\theta_2)$ in coordinates, while the term in $h$ is antisymmetric.} The first non-vanishing order is $\epsilon^2$.
  
  Using $\sqrt{G} \approx \sqrt{G_{U(1)_R}} = w^{-2} r^5$, the following terms may now arise up to order $\varphi^2$ after integrating out the angular directions in the 10d action
  \begin{equation}
   S_\text{5d} = \frac{1}{g_s^2 \kappa_{10}^2} \int \diff^4 x \int \diff r  \, r^5 w(r)^{-2} \left( R_\text{5d}  + \Lambda(r) - \frac{1}{2} \epsilon(r)^2 \, G_\text{5d}^{ij} \partial_i \varphi \partial_j \varphi \right) \,.
  \end{equation}
  Here, we have implemented 5d Lorentz symmetry and used the fact, that there are no linear terms in $\varphi$ once we go to order $\epsilon^2$. Importantly there a no non-derivative terms for $\varphi$, because $\varphi=\text{const.}$ is a point in complex moduli space. Dropping 5d gravity, we arrive  at
  \begin{equation}
   S_\text{5d} = \frac{1}{g_s^2 \kappa_{10}^2} \int \diff^4x \int_{r_\IR}^{r_\UV} \diff r \, \epsilon(r)^2 r^5 \left(-\frac{1}{2} w(r)^{-4} \left(\partial_\mu \varphi\right)^2 - \frac{1}{2} \left(\partial_r \varphi\right)^2 \right) \,.
  \end{equation}
  Explicity calculating $R(G)$ with $\varphi = \varphi(x^\mu,t)$ and expanding to first order in $\abs{z}$ and second order in $\epsilon$ confirms this explicitly.
  
 \subsection{Schr\"odinger Equations and Exact Solutions for Free Fields}
  
  Using the previous subsection and the results of \cite{Hebecker:2015tzo}, we collect the two actions\footnote{For ease of exposition we work with only two real fields although strictly speaking they are both real components of complex fields (see Sect.~\ref{sec:4dSugra}).} that will give us all relevant parameters. Note that we derive the action for $\varphi$ for $c=0$ and that the one for $c$ for $\varphi=0$. We deal with the form of interaction terms in App.~\ref{sec:AxionPeriodicity}.
  \begin{equation}
  \begin{aligned}
   S_\varphi =& \, \frac{1}{g_s^2 \alpha'^4} &&\int \diff^4x \int \frac{\diff r}{r} ~ \epsilon(r)^2 r^6 &&\left(-\frac{1}{2} w(r)^{-4} \left(\partial_\mu \phi_\varphi\right)^2 - \frac{1}{2} \left(\partial_r \phi_\varphi\right)^2 \right) \,,\\
   S_c =& \, \frac{1}{\alpha'^2} &&\int \diff^4x \int \frac{\diff r}{r} ~  r^2 w(r)^{4} &&\left(-\frac{1}{2} w(r)^{-4} \left(\partial_\mu \phi_c\right)^2 - \frac{1}{2} \left(\partial_r \phi_c\right)^2 \right) \,,
   \label{eq:Actions}
  \end{aligned}
  \end{equation}
  where we use a general notation with fields $\phi_i(x^\mu,r)$, $i=\varphi,c$. The equations of motion using the plane wave ansatz $\phi_i(x^\mu,r) = e^{ip_i x} \chi_i(r)$, with $p_i^2 = -m_i^2$, read
  \begin{equation}
  \begin{split}
   w(r)^4 \epsilon(r)^{-2} r^{-5} \partial_r \left( \epsilon(r)^2 r^5 \partial_r \chi_\varphi(r) \right) =& \, -\, m_\varphi^2 \chi_\varphi(r) \,,\\
   r^{-1} \partial_r \left( r w^4 \partial_r \chi_c(r) \right) =& \, - m_c^2 \chi_c(r) \,.
  \end{split}
  \end{equation}
  We insert the warp-factor \eqref{eq:WarpFactor}, dropping logarithmic dependencies, and the $U(1)_R$-breaking factor to arrive at
  \begin{equation}
  \begin{split}
   r^5 \partial_r \left( \frac{1}{r} \partial_r \chi_\varphi(r) \right) =& \, - R^4\, m_\varphi^2 \chi_\varphi(r) \,,\\
   \frac{1}{r} \partial_r \left( r^5 \partial_r \chi_c(r) \right) =& \, - R^4\, m_c^2 \chi_c(r) \,,
  \end{split}
  \end{equation}
  where we introduced the abbreviation $R = \sqrt{g_s M \ap}$ for the normalization of the warp factor of KT and KS. Using the reparametrization
  \begin{equation}
  \begin{aligned}
   x_\varphi =& \frac{R^2 \, m_\varphi}{r} \,, &&&g_\varphi(x_\varphi) &= \frac{1}{r} \chi_\varphi(r(x_\varphi)) \,,\\
   x_c =& \frac{R^2 \, m_c}{r} \,, &&&g_c(x_c) &= r^2 \chi_c(r(x_c)) \,,
  \end{aligned}
  \end{equation}
  we arrive at the following form of the equations of motion
  \begin{equation}
   x_i^2 g_i''(x_i) + x_i g_i'(x_i) + \left(x_i^2 - \alpha_i^2\right) g_i(x_i) = 0 \,,
  \end{equation}
  where $i = \varphi, c$ and where $\alpha_\varphi = 1$ and $\alpha_c = 2$. This is the standard form of the Bessel equation. Plugging back in our reparametrization, we write down the general solutions to the equations of motion
  \begin{equation}
  \begin{split}
   \chi_\varphi(r) =& \, r \left( A \, J_1\left( \frac{R^2 \, m_\varphi}{r} \right) + B \, Y_1\left( \frac{R^2 \, m_\varphi}{r} \right) \right) \,, \\
   \chi_c(r) =& \, \frac{1}{r^2} \left( A \, J_2\left( \frac{R^2 \, m_c}{r} \right) + B \, Y_2\left( \frac{R^2 \, m_c}{r} \right) \right) \,,
  \end{split}
  \end{equation}
  with Bessel functions $J_{\alpha_i}$ and $Y_{\alpha_i}$ of first and second kind respectively.
  
\begin{figure}[ht]
   \centering
   \begin{subfigure}{.4\textwidth}
    \includegraphics[width=1\textwidth]{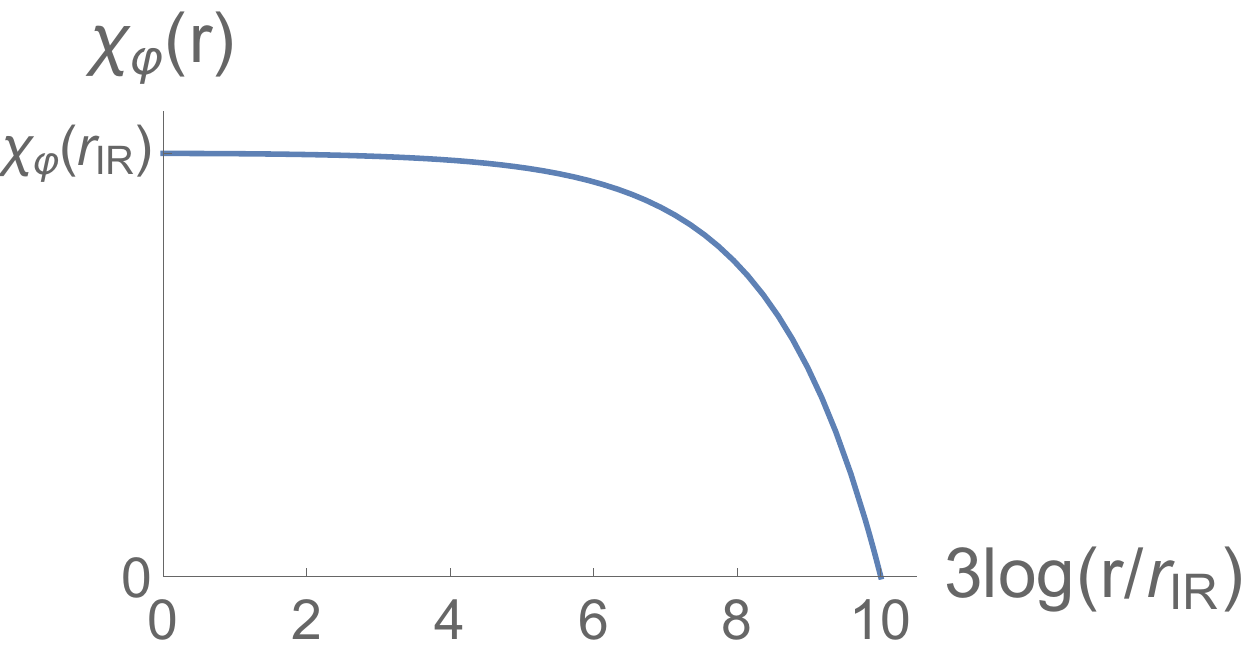}
   \end{subfigure}
   \begin{subfigure}{.4\textwidth}
    \includegraphics[width=1\textwidth]{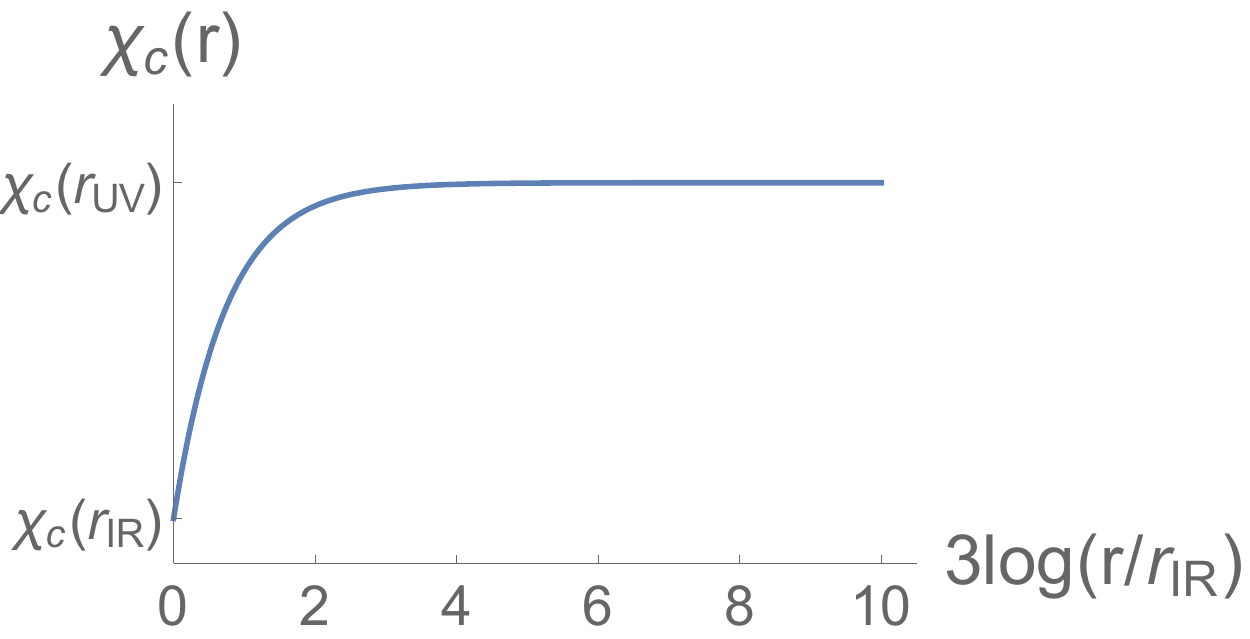}
   \end{subfigure}
   \caption{The lowest lying radial eigenmodes for $\varphi$ and $c$ for $w_\IR \sim 10^{-2}$ plotted in the physical radial coordinate $t = 3 R \log(r/r_\IR)$.}
   \label{fig:ExactSol_Plots}
  \end{figure}

  We now have to apply the boundary conditions
  \begin{equation}\label{eq:boundary-conditions}
  \begin{aligned}
   \chi_\varphi(r_\UV) =& \, 0 \,, &&& \partial_r \chi_\varphi(r_\IR) =& \, 0 \,,\\
   \partial_r \chi_c(r_\UV) =& \, 0 \,, &&& w_\IR \partial_r \chi_c(r_\IR) =& \, \frac{\delta}{R} \chi_c(r_\IR) \,,
  \end{aligned}
  \end{equation}
  for some $\calO(1)$-number $\delta$ depending on how we model the flux distribution in the IR. While the UV boundary conditions are easily implemented to fix the constants $A$ and $B$ up to an overall normalization, the ansatz $m_i \propto \frac{w_\IR^2}{R}$ gives consistent solutions to the IR boundary conditions when using the small argument expansions of the Bessel functions
  \begin{equation}
  \begin{split}
   m_\varphi^2 =& \, 8 \frac{w_\IR^4}{R^2} \,,\\
   m_c^2 =& \, \frac{8 \delta}{4 + \delta} \frac{w_\IR^4}{R^2} \,,
   \label{eq:Eigenmasses}
  \end{split}
  \end{equation}
  where we now normalized the warp factor $w_\UV = 1$. We plot the solutions with $\delta =1$, $w_\IR = e^{-10/3} \sim 10^{-2}$ and $m_i$ as above in Fig.~\ref{fig:ExactSol_Plots}.
  
\begin{figure}[ht]
   \centering
   \begin{subfigure}{.4\textwidth}
    \includegraphics[width=1\textwidth]{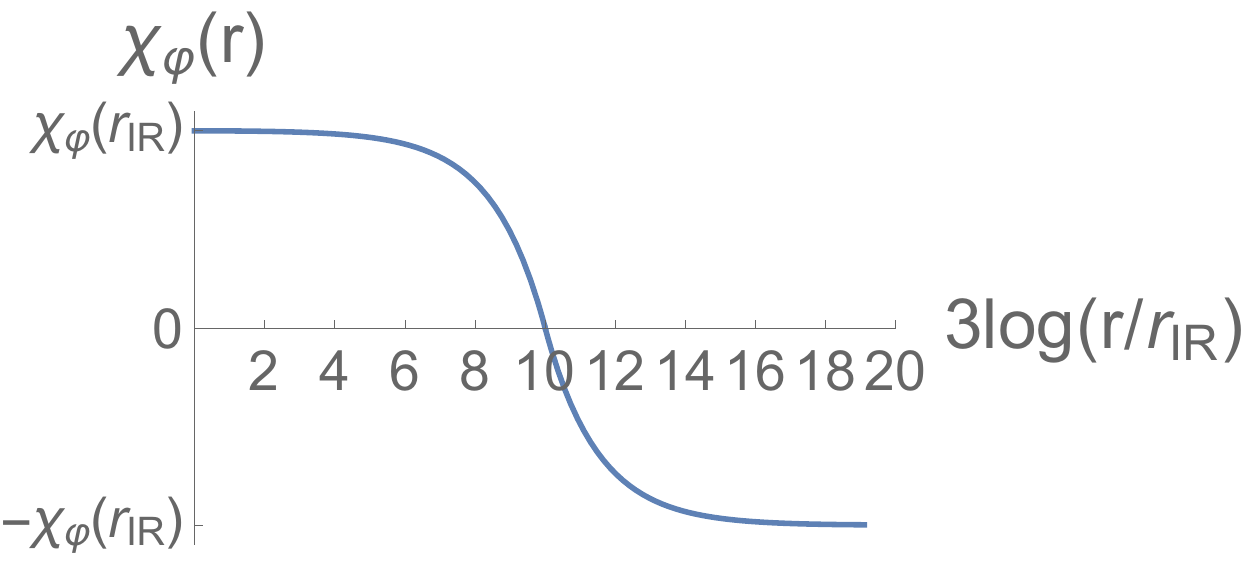}
   \end{subfigure}
   \begin{subfigure}{.4\textwidth}
    \includegraphics[width=1\textwidth]{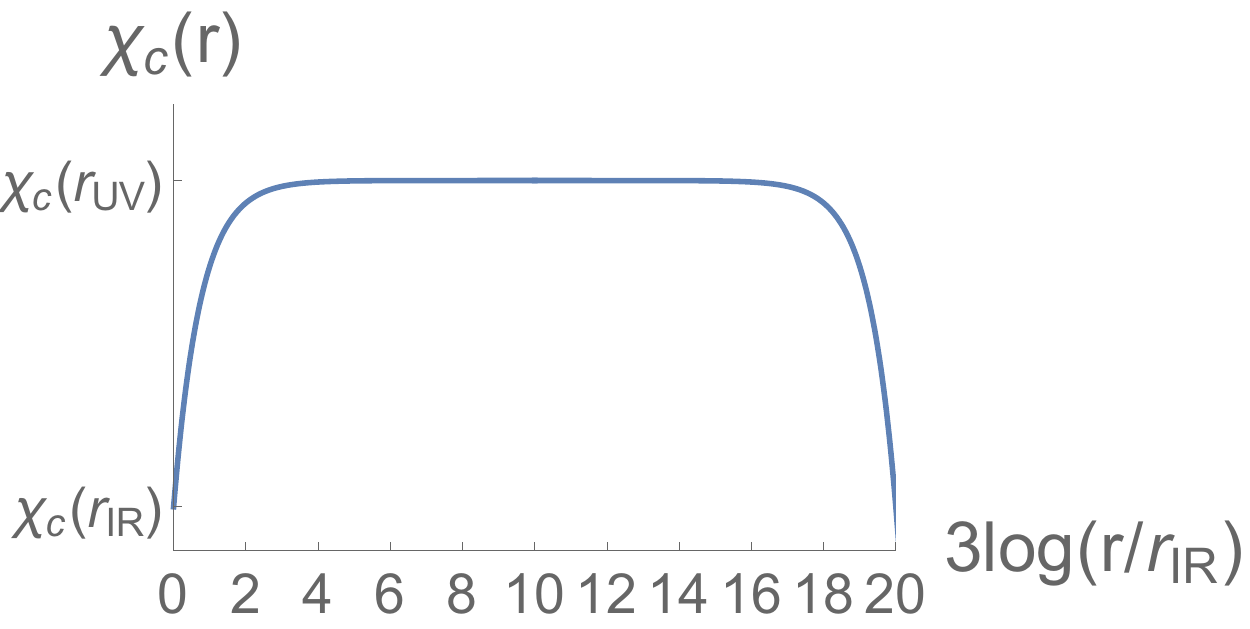}
   \end{subfigure}
   \caption{The eigenmodes for $\varphi$ and $c$ on the double throat. Here, $t<10$ corresponds to the first throat, $t>10$ corresponds to the second, with $t=0$ and $t=10$ corresponding to the respective ends of the throats.}
   \label{fig:ExactSol_PlotsBifid}
  \end{figure}

  Referring back to the discussion in Sect.~\ref{sec:CYBreakingPotential}, we plot the expected solutions over the entirety of the double throat in Fig.~\ref{fig:ExactSol_PlotsBifid}, where we simply mirrored one throat at the UV end with appropriate symmetry, so both $r=r_\IR$ and $r=2 r_\UV - r_\IR$ correspond to IR regions.
  
  We are left with calculating the 4d kinetic terms. From Fig.~\ref{fig:ExactSol_Plots}, we would guess (correctly) that $\varphi$ is an IR mode, while $c$ is a UV mode. We derive this by considering the lowest KK mode $\phi_0(x^\mu,r) = f_0(x^\mu) \chi_0(r)$, where $f_0$ is now an arbitrary superposition of plane waves, to find the kinetic terms induced by the actions \eqref{eq:Actions}
  \begin{equation}
  \begin{aligned}
   &f_\varphi^2 = \,  \, \frac{1}{g_s^2 \alpha'^4} \, \int_{r_\IR}^{r_\UV} \frac{\diff r}{r} \, \epsilon(r)^2 \, r^6 \, w(r)^{-4} \chi_\varphi(r)^2 \approx \, \frac{R^6}{g_s^2\alpha'^4} w_\IR^2\,,\\
   &f_c^2 = \,  \frac{1}{\alpha'^2} \, \int_{r_\IR}^{r_\UV} \frac{\diff r}{r} \, r^2 \,  \chi_c(r)^2 \approx \, \frac{R^2}{\alpha'^2}\, .
   \label{eq:KineticTerms}
  \end{aligned}
  \end{equation}
  Here, we have again expanded the solution in $w_\IR \ll 1$, and neglected logarithmic corrections. Furthermore we have normalized the internal field profiles such that $\chi_c(r_\UV)=\chi_{\varphi}(r_\IR)=1$.
  
  Comparing with the quadratic terms in $c$ in the potential \eqref{eq:fluxpotential} and the quadratic terms in $\varphi_i$ in the potential \eqref{eq:CYBreakingPot}, we find the parameters $\mu$ and $\Lambda$
  \begin{equation}
   \mu^4 = m_c^2 \, f_c^2 \propto w_\IR^4 \, , \quad \Lambda^4 = m_\varphi^2 \, f_\varphi^2 \propto w_\IR^6 \,.
  \end{equation}

 \section{Axion Decay Constant and Potential Parameters}
 \label{Appendix:AxionDecayConstant}
 
  In this appendix we extend the results of the previous App.~\ref{sec:ExactResults} by including logarithmic corrections to the first-order approximations and by taking care of the different length scales that may appear in the parameters \eqref{eq:coefficients} of the effective potential \eqref{eq:vc}. 
  
  We compute the axion decay constant (without accounting for monodromy factors) for the $c$-axion from the 10d action, paying attention to numerical factors of $\mathcal{O}(1)$. We may dimensionally reduce the $|F_3|^2$ term of the 10d type IIB SUGRA and plug in $C_2=2\pi \alpha' c(x) \omega_{\Sigma}$, where $\omega_{\Sigma}$ is the associated (quasi-)harmonic form. This leads to
  \begin{equation}
   \frac{1}{(2\pi)^7\alpha'^4}\left(-\frac{1}{2}\int *|F_3|^2\right)=\int \diff^4 x \left(-\frac{1}{2}f^2(\del c)^2\right)\, ,
  \end{equation}
  with
  \begin{equation}\label{eq:axion-decay-constant}
   f^2=(2\pi)^2 g_s^2 \frac{\int_{\CY}\diff^6 y\,  \sqrt{g_{6}}\,w^2\,\alpha'^2|\omega_{\Sigma}|^2}{\int_{\CY}\diff^6 y\,\sqrt{g_6}\,w^2}M_\Pl^2\equiv (2\pi)^2g_s^2 \frac{\alpha'^2}{(\text{Vol}(S^2)|_{\UV})^2}M_\Pl^2\, ,
  \end{equation}
  where $g_s$ is the string coupling, $g_6$ is the internal 6d metric, $w^2$ is the warp factor and $M_\Pl$ is the four-dimensional Planck mass. Due to the appearance of the warp factor, the integrals are dominated by UV contributions. Thus, for large values of the overall K\"ahler modulus $\text{Re }T\gg N_\text{D3}$ we have $f^2\sim g_s (\text{Re }T)^{-1}$ in Planck units\footnote{Recall that $\text{Re }T$ measures a $4$-cycle volume in 10d Einstein frame.}, where $N_\text{D3}$ is the total D$3$ brane charge stored in fluxes (and mobile D$3$ branes). This is consistent with the 4d SUGRA of Sect.~\ref{sec:4dSugra}. We expect to be allowed to make the string frame volume as small as  $\mathcal{O}((g_sN_\text{D3})^{3/2})$. At even smaller volume backreaction of fluxes becomes significant throughout the CY and we lose the notion of an unwarped bulk CY \cite{Giddings:2001yu}. For simplicity we now take into consideration only the D$3$ brane charge stored in a single throat, i.e. $N_\text{D3}=MK$. In this regime the top of the throat marginally fits into the bulk CY and we expect the ratio of integrals in \eqref{eq:axion-decay-constant} to be well approximated by the value of $\alpha'^2|\omega_{\Sigma}|^2$ at the top of the KS throat. Since the hierarchy is set by $\log w_{\IR}^{-1}\sim \frac{K}{g_sM}$ it follows immediately that $f$ scales as $f\sim \sqrt{\frac{g_s}{MK}}M_\Pl\sim \frac{1}{M}\log(w_{\IR}^{-1})^{-1/2}M_\Pl$.
  
  We can compute this in more detail by using the asymptotic form of the throat metric far away from the deformation, i.e.~the KT metric \eqref{eq:KTMetric}
  \begin{align}
   \diff s_6^2&=w^{-2}(\diff r^2+r^2 \diff s^2_{T^{1,1}})\, ,\quad   \diff s^2_{T^{1,1}}=\frac{1}{9}(g^5)^2+\frac{1}{6}\sum_{i=1}^{4}(g^i)^2\, ,\\ w^2&=\frac{2\sqrt{2}}{9}\frac{r^2}{g_sM\alpha'}\log(r/r_{\IR})^{-\frac{1}{2}}\, ,\quad \omega_{\Sigma}=\frac{1}{8\pi}(g^1\wedge g^2+g^3\wedge g^4)\, .
  \end{align}
  We see, that the physical radius at the top of the throat is given by
  \begin{equation}
   R_\text{throat}^2 = \frac{9}{2\sqrt{2}} g_s M \ap \sqrt{\log{r_\UV/r_\IR}} \approx \frac{9}{2\sqrt{2}} g_s M \ap \sqrt{\log{w_\IR^{-1}}} \,,
  \end{equation}
  where in the last step we use the normalized warp factor with $w_\UV = 1$. Taking into account angular dependencies of $\T$, we arrive at $\alpha'|\omega_{\Sigma}|=(3\pi g_s M)^{-1}\log(w_{\IR}^{-1})^{-\frac{1}{2}}$, and hence
  \begin{equation}\label{eq:axiondecayconstant}
   f\approx \frac{2}{3M} \log(w_{\IR}^{-1})^{-1/2}M_\Pl\, .
  \end{equation}
  
  Actually, we are interested in the case where two or more throats each store only part of the total D$3$ brane charge and also fluxes on other cycles contribute to $N_\text{D3}$. It is useful to define as $r_\text{D3}^i$ the fraction of the total D$3$ brane charge that is stored in the $i$-th throat, i.e.
  \begin{equation}
  r_\text{D3}^i\equiv M_i K^i/N_\text{D3}\, ,\quad i=1,...,n\, .
  \end{equation}
  Since the decay constant scales as $(\text{Re }T)^{-1/2}$ and the marginal case corresponds to $\text{Re }T\sim  N_\text{D3} = M_i K^i /r_\text{D3}^i$, for any $i=1,...,n$, we should really replace \eqref{eq:axiondecayconstant} by
  \begin{equation}
  f\approx  \frac{2}{3M_i}(r_\text{D3}^i)^{1/2}\log((w_{\IR}^i)^{-1})^{-1/2}M_\Pl\, ,
  \end{equation}
  and this expression is correct when evaluated for any $i=1,...,n$.
  
  In the double throat case, we have $r_\text{D3}^1=r_\text{D3}^2\equiv r_\text{D3}$ and the effective (monodromy-) potential \eqref{eq:vc} reads
  \begin{equation}
   V(c)/M_\Pl^4 \sim w_\IR^6 \left(1-\cos\left(\frac{3}{(r_\text{D3})^{1/2}} \sqrt{\log(1/w_\IR)} \frac{c}{M_\Pl}\right)\right) \,.
  \end{equation}
  Even in this marginal case, the decay constant of $c$ is slightly sub-Planckian in regimes of parametric control since then $w_{\IR}\ll 1$, and $r_\text{D3}<1$.
  
  We may also derive the quantities given in \eqref{eq:coefficients}: Using \eqref{eq:Actions} the kinetic term of $\varphi$ is given by
  \begin{equation}
  f_{\varphi}^2\sim \frac{(g_sM\alpha')^6}{g_s^2 \alpha'^4} w_\IR^{2} \sim \frac{(g_sM)^3}{\mathcal{V}}w_\IR^2 M_\Pl^2\, ,
  \end{equation}
  where we use that the integral for the kinetic term \eqref{eq:KineticTerms} is dominated by the IR, such that we may insert the IR throat radius, $R^2 \sim g_s M \ap$, and where we have made use of the fact that $M_\Pl^2\sim g_s^{-2}\alpha'^{-1}\mathcal{V}$, with the (bulk) CY volume $\mathcal{V}$ as measured in string units. Neglecting geometric back-reaction of the double throat system in the UV, the flux energy density $M^2 \mu^4$ can be understood as a pure IR effect: It is induced by an excursion of $\varphi_{1,2}$ in the IR whilst keeping $c$ fixed (for the factor of $M^2$ compare \eqref{eq:fluxpotential}). Therefore the resulting mass must be of the order of the warped KK-scale $m_\text{wKK}^2 \sim R^{-2} w_\IR^2 \sim (g_sM\alpha')^{-1}w_\IR^2$
  \begin{equation}
  \mu^4\sim \frac{1}{M^2}f_{\varphi}^2m_\text{wKK}^2\sim \frac{g_s^4}{ \mathcal{V}^2}w_{\IR}^4M_\Pl^4\, .
  \end{equation}
  Finally, the scale $\Lambda^4$ is dominated by contributions from the bulk CY (which we again assume to have only one length-scale, $ R_{\CY}^2\sim \mathcal{V}^{1/3}\alpha'$), so
  \begin{align}
  \Lambda^4\sim &\underbrace{\frac{1}{g_s^2}\frac{1}{\alpha'^4}}_{\sim M_{\Pl,\text{10d}}^8}\quad \underbrace{\mathcal{V}\alpha'^3}_{\sim \int \diff^6y\sqrt{g_{\CY}}}\quad  \underbrace{\epsilon^2(r_{\UV})}_{\sim \text{UV tail of deformation}}\quad  \underbrace{\frac{1}{\mathcal{V}^{1/3}\alpha'}}_{\sim \mathcal{L}_\text{10d}  \supset (\nabla \varphi)^2 \propto 1/R_\text{CY}^2} \nonumber\\
  \sim & \, \frac{g_s^2}{\mathcal{V}^{4/3}}\epsilon^2(r_{\UV})M_\Pl^4\sim \frac{g_s^2}{\mathcal{V}^{4/3}}\log(w_\IR^{-1})^{-3/2}w_\IR^6M_\Pl^4\, .
  \end{align}
  Here, we have made use of the fact that $\epsilon^2(r_\UV)\sim (r_\IR/r_\UV)^6\sim w_{\IR}^6\log(w_\IR^{-1})^{-3/2}$. The scale $\Lambda^4$ determines the axion potential and can be compared with the SUGRA result. Using the K\"ahler potential \eqref{eq:Kahlerpot} one easily shows that the parametric dependencies come out correctly.
  
  Going to the limit where the throats marginally fit into the bulk CY means taking $\mathcal{V}\sim (g_sN_\text{D3})^{3/2}\sim (r_\text{D3})^{3/2}(g_sM)^3\log(w_\IR^{-1})^{3/2}$. In this limit,
  \begin{align}
  \Lambda^4/M_{\Pl}^4&\sim (r_\text{D3})^{-2}\frac{g_s^2}{(g_sM)^4}\log(w_\IR^{-1})^{-7/2}w_\IR^6\, ,\quad \mu^4/M_\Pl^4\sim (r_\text{D3})^{-3}\frac{g_s^4}{(g_sM)^{6}}\log(w_\IR^{-1})^{-3}w_\IR^4\, ,\nonumber\\
  f_{\varphi}^2/M_\Pl^2&\sim (r_\text{D3})^{-3/2}\log(w_\IR^{-1})^{-3/2}w_\IR^2\, .
  \end{align}
  For the double throat case considered in Sect.~\ref{sec:Upshot}, we may assume a configuration where almost all the D$3$ brane charge resides in the two throats, and therefore $r_\text{D3}\approx 1/2$.
 \section{Background on Multi Conifolds}\label{Appendix:BackgroundOnConifolds}
  
  In this appendix we discuss preliminaries that are important for Sect.~\ref{sec:4dSugra}. We follow mainly Chapter $8$ of \cite{Greene:1996cy}.
  
  For a general CY threefold $M$ we can choose $2h^{2,1}+2$ three-cycles $\mathcal{A}^a$, $\mathcal{B}_a$, $a=1,...,h^{2,1}+1$ as a symplectic basis of $H^3(M)$, i.e.
  \begin{equation}
   \int_{\mathcal{A}^b}\alpha^a=\int_M \alpha^a\wedge \beta_b=\delta^a_b\, ,\quad \int_{\mathcal{B}_b}\beta_a=\int_M \beta_a\wedge \alpha^b=-\delta_a^b\, ,
  \end{equation}
  where $\alpha^a$ and $\beta_a$ are the harmonic three-forms that are Poincaré dual to $\mathcal{B}_a$, respectively $\mathcal{A}^a$. One may define the periods
  \begin{equation}
   z_a=\int_{\mathcal{A}^a}\Omega\, ,\quad G^a=\int_{\mathcal{B}_a}\Omega\, ,
  \end{equation}
  where $\Omega$ is the holomorphic three-form. The $z_a$ form a set of projective coordinates on complex structure moduli space and the $G^a$ are functions of them. We are interested in what happens when $n$ cycles $\gamma^i$ with $m$ homology relations among them shrink at a conifold point in moduli space. The Picard-Lefschetz formula states that upon encircling a conifold point in moduli space, a three-cycle $\delta$ undergoes the monodromy \cite{picard1897theorie,MR0033557,Greene:1995hu}
  \begin{equation}
   \delta \longrightarrow \delta + \sum_{i=1}^n (\delta \cap \gamma^i)\gamma^i\, .
   \label{eq:PicardLefschetz}
  \end{equation}
  Knowing this monodromy transformation is enough to determine that
  \begin{equation}
  \int_{\delta}\Omega=\frac{1}{2\pi i}\sum_{i=1}^{n}(\delta \cap \gamma^i)\int_{\gamma^i}\Omega\, \log(\int_{\gamma^i}\Omega)+ \text{single-valued}\, .
  \end{equation}
  We may choose $n-m$ of the degenerating cycles as part of the basis $\mathcal{A}^i=\gamma^i$ for $i=1, \ldots,n-m$, while the remaining $m$ are integer linear combinations $\gamma^{i}=\sum_{a=1}^{n-m} c^i_a \mathcal{A}^a$ for $i=n-m+1, \ldots, n$. By applying \eqref{eq:PicardLefschetz} to the cycles $\mathcal{B}_a$ we arrive at
  \begin{equation}
   G^a=\int_{\mathcal{B}_a}\Omega=\frac{1}{2\pi i}z_a \log(z_a)+\frac{1}{2\pi i}\sum_{i=n-m+1}^n c^i_a z_i \log(z_i)+g^a(z)\, , \quad a=1,...,n-m\, ,
  \end{equation}
  where $g^a(z)$ are $n-m$ holomorphic functions. We have defined $z_i\equiv \sum_{a=1}^{n-m}c^i_a z_a$ for $i=n-m+1,\ldots,n$, i.e.~$z_i \equiv \int_{\gamma^i} \Omega$ when applying \eqref{eq:PicardLefschetz}\footnote{When using a local expression for the holomorphic three-form $\Omega$ in the vicinity of smoothed conical singularity described by \eqref{eq:DeformedConifold} one can calculate $\int_{\gamma^i} \Omega = z_i$ \cite{Cachazo:2001jy}. This identifies the $z_i$ defined here with the local deformation parameter of the $i$-th throat.}. At frozen values of $z^a$, $a=n-m+1,...,h^{2,1}+1$ the periods associated to other cycles, $G^a=\int_{\mathcal{B}_a}\Omega$, with $a=n-m+1,...,h^{2,1}+1$, are holomorphic in the complex structures that parametrize the multi conifold deformations, i.e.~in the $z^i$, with $i=1,...,n-m$. In what follows we denote by $z^a$ only the multi conifold deformation parameters.
  
  We may now evaluate the GVW superpotential $W=\int_M G_3\wedge \Omega$ where we choose flux quanta $M_a$ and $K^a$ according to $G_3=-\sum_{a=1}^{n-m}(M_a\alpha^a-\tau K^a \beta_a)$. Using that
  \begin{equation}
   \int_M \alpha^a\wedge \Omega=-\int_{\mathcal{B}_a}\Omega=-G^a\, ,\quad \int_M \beta_a\wedge \Omega=-\int_{\mathcal{A}^a}\Omega=-z_a\, ,
  \end{equation}
  one obtains
  \begin{align}
   W(z_a)=&\sum_{a=1}^{n-m}\frac{M_a}{2\pi i} z_a\log(z_a)+ \sum_{i=n-m+1}^{n}\frac{M_i}{2\pi i}z_i\log (z_i)\nonumber\\
   &+\sum_{a=1}^{n-m}M_a g^a(z) -\tau \sum_{a=1}^{n-m}K^a z_a +\hat{\hat{W}}_0(z_a)\, , 
  \end{align}
  where we have \textit{defined} $M_i\equiv \sum_{a=1}^{n-m}c^i_a \, M_a$, and the holomorphic function $\hat{\hat{W}}_0(z_a)$\linebreak parametrizes the contributions from fluxes on other cycles. We may use the $z_a$ and $z_i$ with $i=n-m+1,...,n$ on the same footing by interpreting our definition of the $z_i$ as $m$ constraint equations
  \begin{equation}
   0=P^I\equiv \sum_{i=1}^{n}p^I_i z_i \equiv  z_{n-m+I}-\sum_{a=1}^{n-m}c^{n-m+I}_a z_a \, ,\quad I=1,...,m\, .
  \end{equation}
  Here, the $m\times n$ matrix $p^I_i$ is implicitely defined as
  \begin{equation}
   p^I_i = \begin{cases} -c_i^{n-m+I}  \, ,\, &i=1,\ldots,n-m \,,\\ \delta_{i}^{n-m+I}  \, ,\, &i=n-m+1 ,\ldots,n\,.\end{cases}
  \end{equation}
  We may now write the superpotential as
  \begin{equation}\label{eq:n-conifold-superpotential}
   W(z_i)=\sum_{i=1}^{n}\left(M_i\frac{z_i}{2\pi i} \log(z_i)+M_ig^i(z)-\tau K^i z_i\right)+\sum_{I=1}^{m}\lambda_I P^I+\hat{\hat{W}}_0(z_i)\, ,
  \end{equation}
  with $m$ Lagrange multipliers $\lambda_I$. The homology relation is now enforced via the F-term of the fields $\lambda_I$, compare \cite{Cachazo:2001jy} where Lagrange multipliers in the superpotential are also used to impose additional constraints on chiral superfields. In doing so, we have defined $g^i$ to be zero for $i>n-m$. 
  
  The $F_3$-flux on $\gamma^i$ is given by $M_i$. By the definition of $M_i$ for $i=n-m+1,\ldots,n$, the flux numbers automatically fulfill $\sum_{i=1}^n p^I_i M_i=0$ for all $I$. In democratic terms, the $n$ flux numbers $M_i$ must be chosen in compliance with the $m$ homology constraints $\sum_{i=1}^n p^I_i M_i=0$. The $H_3$-flux on $\mathcal{B}_a$ is given by $K^a + \sum_{I=1}^m c^{n-m+I}_a K^{n-m+I}$, as this is the coefficient appearing in front of $z_a$. In other words the $n-m$ flux quantization conditions read $K^a + \sum_{I=1}^m c^{n-m+I}_a K^{n-m+I} \in \Z$. Note that we may transform $K^i\to K^i+\sum_I \alpha_I p^I_i $ for any $\alpha\in\C^m$ because the superpotential is left invariant upon imposing the constraint equations, that is to say, we can undo such a transformation by also shifting the Lagrange multipliers $\lambda^I \to \lambda^I + \tau \alpha^I$. Of course, the flux quantization conditions are invariant under these shifts. Finally, the K\"ahler potential is given by
  \begin{align}\label{eq:Kahlerpot_cs}
  K_\text{cs}(z_i,\bar{z}_i)=&-\log\left(-i \int \Omega \wedge \bar{\Omega} \right)=-\log\left(ig_K(z)-i\overline{g_K}(z) + \sum_{a=1}^{n-m}i\bar{z_a}G^a(z)+c.c.\right)\\
  =&-\log\left(ig_K(z)-i\overline{g_K(z)}+\sum_{i=1}^{n}\left[\frac{|z_i|^2}{2\pi}\log(|z_i|^2)+i\bar{z}_ig^i(z)-iz_i\overline{g^i(z)}\right]\right)\, ,
  \end{align}
  where $g_K=\sum_{a=n-m+1}^{h^{2,1}+1}\overline{z_a} G^a(z)$ encodes contributions from other cycles. It is holomorphic in $z_a$, $a=1,...,n-m$. Note that despite the democratic formulation, the K\"ahler and superpotential are strictly defined only along complex structure moduli space, where $P^I=0$. As explained in Sect.~\ref{sec:4dSugra} we propose to extend the domain of these functions to the deformation space parametrized by \textit{all} $z_i$ by introducing general Taylor expansions of $g^i(z_i)$, $g_K(z_i)$ and $\hat{\hat{W}}_0(z_i)$ in \eqref{eq:thresholdcoeff}.

   \addcontentsline{toc}{section}{References}
  
 \bibliographystyle{JHEP}
 \bibliography{references}

\end{document}